\documentclass[cits]{PoS}

\usepackage{graphicx}
\usepackage{amssymb}
\usepackage{amsmath}

\makeatletter
\def\simleq{\mathrel{\mathpalette\gl@align<}}
\def\simgeq{\mathrel{\mathpalette\gl@align>}}
\def\gl@align#1#2{\lower.6ex\vbox{\baselineskip\z@skip\lineskip\z@
     \ialign{$\m@th#1\hfill##\hfil$\crcr#2\crcr\sim\crcr}}}
\makeatother

\newcommand{\Pu}{p_{\uparrow}}

\newcommand{\Nu}{n_{\uparrow}}
\newcommand{\Nd}{n_{\downarrow}}

\title{Nuclear physics from lattice simulations}

\ShortTitle{Nuclear physics from lattice simulations}

\author{\speaker{Takumi Doi}\\%
Theoretical Research Division, Nishina Center, RIKEN, Wako 351-0198, Japan\\
E-mail: \email{doi@ribf.riken.jp}}

\author{for HAL QCD Collaboration}

\author{\includegraphics[width=.30\textwidth]{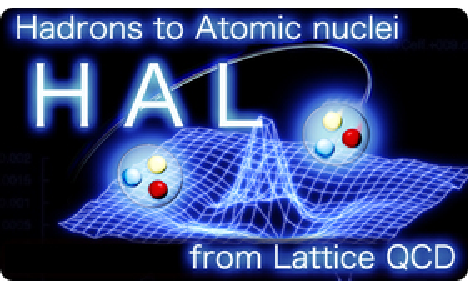}}

\abstract{
We review recent lattice QCD activities 
with emphasis on the impact on nuclear physics.
In particular, the progress toward the 
determination of nuclear and baryonic forces (potentials)
using Nambu-Bethe-Salpeter (NBS) wave functions is presented.
We discuss major challenges for multi-baryon systems on the lattice:
(i)  signal to noise issue and
(ii) computational cost issue.
We argue that the former issue can be avoided by extracting 
energy-independent (non-local) potentials 
from time-dependent NBS wave functions
without relying on the ground state saturation,
and the latter cost is drastically reduced by developing a novel ``unified contraction algorithm.''
The lattice QCD results for nuclear forces, hyperon forces and three-nucleon forces are presented,
and physical insights are discussed.
Comparison to results from the traditional L\"uscher's method
is given, and open issues to be resolved are addressed as well.
}

\FullConference{The 30 International Symposium on Lattice Field Theory - Lattice 2012,\\
		June 24-29, 2012\\
		Cairns, Australia}

\begin{document}

\section{Introduction}
\label{sec:intro}

One of the most fundamental question in nuclear physics is 
how nuclear physics itself emerges from the underlying theory, 
Quantum Chromodynamics (QCD).
To answer this question, 
nuclear (and hadronic) forces are expected to play a vital role.
They correspond to a particular representation of hadronic S-matrices,
or scattering phase shifts,
and serve as the most fundamental quantities
in nuclear physics.
Traditionally, nuclear potentials are phenomenologically determined
using experimental two-nucleon (2N) scattering phase shifts as inputs.
Recent ``realistic two-nuclear forces (2NF)'' can reproduce several thousands of
2N scattering data with $\chi^2/{\rm dof} \sim 1$ with 30-40
fitting parameters.
The major challenge imposed on nuclear and particle physics is 
to determine hadronic potentials 
in the first-principles method of QCD, such as lattice QCD simulations.

Once hadronic forces are determined directly from QCD,
various applications to nuclear and astrophysical phenomena are in order.
Some examples include the structures and reactions of atomic nuclei,
and 
the equation of state (EoS) of nuclear matter.
The latter is relevant not only to 
nuclear saturation
but also to the physics of neutron stars and supernova explosions.
Note, however, that not only 2NF but also other hadronic forces
play an important role.
For instance, detailed information of hyperon-nucleon (YN) and
hyperon-hyperon (YY) interactions is necessary to understand 
the structure of neutron stars with hyperon core.
They are essential inputs to the strangeness physics conducted in
the J-PARC experiments as well.
Three-nucleon forces (3NF) are other quantities which have vital roles 
in modern nuclear physics and astrophysics.
They have an impact on binding energies of light nuclei, nuclear EoS
and the maximum mass of neutron stars. 
Recently, the effect of 3NF on properties of neutron-rich nuclei,
which would be relevant to supernova nucleosynthesis, 
is also recognized.
Since experimental information for YN, YY forces and 3NF are quite limited,
lattice QCD predictions are highly awaited.

Traditional approach to obtain hadronic interactions in lattice QCD
is the L\"uscher's finite volume method~\cite{Luscher:1990ux}.
It can connect the energy of the two-particle system in a finite lattice box
to the elastic scattering phase shift. 
Lattice QCD studies for nuclear physics in this approach are given 
in Refs.~\cite{Fukugita:1994ve,Beane:2006mx,Beane:2009gs,Beane:2009py,Yamazaki:2009ua,Yamazaki:2011nd,Yamazaki:2012hi,
Beane:2010em,Beane:2010hg,Beane:2011iw,Beane:2012ey,Beane:2012vq}.
In principle, it is also possible to calculate the tower of 2N energy spectra on the lattice
and obtain the phase shifts at the corresponding energies,
which may be converted to nuclear forces. 
In practice, however, the computations of energy spectra are 
usually limited only for a few excited states at most,
making potentials out of reach of lattice simulations.
One may consider the effective field theory (EFT) approach to ease this problem,
while the results could suffer from uncertainties 
of 
EFT in baryon sector.

To overcome this problem, 
a new approach to hadronic interactions in lattice QCD, now called the HAL QCD method,
has been proposed recently~\cite{Ishii:2006ec,Aoki:2009ji}.
Utilizing the Nambu-Bethe-Salpeter (NBS) wave function on the lattice,
a potential is extracted through the corresponding Schr\"odinger equation.
Since the information of phase shift is embedded in the NBS wave function
at the asymptotic (non-interacting) region,
it is guaranteed that the obtained potential is faithful
to the phase shift by construction.
Resultant (parity-even) 2NF in this approach
are found to have desirable features
such as
attractive wells at long and medium
distances  and central repulsive cores at short distance.
The method has been successfully applied to 
more general hadronic interactions%
~\cite{Nemura:2008sp,Inoue:2010hs,Doi:2010yh,Sasaki:2010bi,Inoue:2010es,Aoki:2011gt,Ikeda:2011qm,Murano:2011nz,Inoue:2011ai,Murano:2011aa,HALQCD:2012aa,Doi:2011gq,Charron:lat2012,Kurth:lat2012,Ishii:lat2012}.
See Refs.~\cite{Aoki:2009ji,Aoki:2012tk} for recent reviews.

Up to now, lattice simulations for multi-baryon systems 
have been carried out at rather heavy quark masses,
while ultimate objective is to perform simulations at the physical point
with infinite volume extrapolation and continuum extrapolation.
Proceeding toward this goal, we have to meet two major challenges.
The first one is the so-called signal to noise (S/N) issue.
Actually, it is well known that S/N is ruined exponentially
for lighter pion mass, larger volume and/or larger baryon number in the system~\cite{Lepage:1989hd}.
The second challenge is the computational cost of the multi-baryon correlators,
namely, the cost of the contractions~\cite{Doi:2012xd}.
Since the cost grows factorially $\times$ exponentially for a larger baryon number $A$ in the system,
it becomes enormous for systems with $ A > 2 $.
In this report, we present a recent breakthrough for each of these issues~\cite{HALQCD:2012aa,Doi:2012xd}.

This report is organized as follows.
We first give a brief review of the HAL QCD method in Sec.~\ref{sec:formalism}.
In Sec.~\ref{sec:challenges}, we explain the major challenges of 
(i) S/N issue and (ii) computational cost issue in lattice simulations for multi-baryon systems,
and recent breakthroughs for these issues are given.
We present the results of lattice numerical simulations 
for nuclear forces (Sec.~\ref{sec:2NF}),
hyperon forces (Sec.~\ref{sec:YN}) and three-nucleon forces (Sec.~\ref{sec:3NF}).
In Sec.~\ref{sec:luscher},
we review the results from the traditional L\"uscher's method.
Comparisons between different groups/approaches are given,
and open issues to be resolved are addressed.
Sec.~\ref{sec:summary} is devoted to conclusions and outlook.

\vspace*{-1mm}
\section{Formalism}
\label{sec:formalism}
\vspace*{-1mm}

We explain the HAL QCD method by considering the 2N potential as an illustration.
We consider the (equal-time) NBS wave function in the center-of-mass frame,
\begin{eqnarray}
\phi_{2N}^{W}(\vec{r}) \equiv \langle 0 | N(\vec{r},0) N(\vec{0},0) | 2N, W \rangle_{\rm in} ,
\end{eqnarray}
where 
$N$ is the nucleon operator and
$|2N, W \rangle_{\rm in}$ denotes the asymptotic in-state of the 2N system 
at the total energy of $W = 2\sqrt{k^2+m_N^2}$
with the nucleon mass $m_N$ and the relative momentum $k \equiv |\vec{k}|$.
For simplicity,
we omit other quantum numbers such as spinor/flavor indices.
For the purpose of clarification, 
we here consider the elastic region,  $W < W_{\rm th} = 2m_N + m_\pi$,
while the method can be extended above inelastic threshold~\cite{Aoki:2011gt,Aoki:E-indep:inelastic}.
The most important property of the NBS wave function is that
it has a desirable asymptotic behavior%
~\cite{Luscher:1990ux, Aoki:2009ji, Lin:2001ek, Aoki:2005uf, Ishizuka:2009bx},
%
\begin{eqnarray}
\phi_{2N}^W (\vec{r}) \propto \frac{\sin(kr-l\pi/2 + \delta_l^W)}{kr}, 
\quad 
r \equiv |\vec{r}| \rightarrow \infty,
\end{eqnarray}
where 
$\delta_l^W$ is the scattering phase shift
with the orbital angular momentum $l$.
Exploiting this feature,
we define the (non-local) 2N potential, $U_{2N}(\vec{r},\vec{r}')$,
through the following Schr\"odinger equation,
\begin{eqnarray}
%
H_0 \phi_{2N}^W(\vec{r})
+ \int d\vec{r}' U_{2N}(\vec{r},\vec{r}') \phi_{2N}^W(\vec{r}')
= E_{2N}^W \phi_{2N}^W(\vec{r}) ,
\label{eq:Sch_2N:tindep}
\end{eqnarray}
where 
$H_0 = -\nabla^2/(2\mu)$ and
$E_{2N}^W = k^2/(2\mu)$ with the reduced mass $\mu = m_N/2$.
It is evident that
$U_{2N}$ defined in this way 
is faithful to the phase shift by construction.

Another important property is that, 
while $U_{2N}$ could be energy-dependent in general,
it is possible to construct $U_{2N}$
so that it becomes energy-independent~\cite{Aoki:2009ji,Aoki:2012tk,Aoki:E-indep:inelastic}.
The outline of the proof can be given as follows.
We first introduce a norm kernel 
${\cal N}_{W_i,W_j} \equiv \int d\vec{r} \overline{\phi_{2N}^{W_i}(\vec{r})} \phi_{2N}^{W_j}(\vec{r})$
for $W_i, W_j < W_{\rm th}$,
and define ${\cal N}^{-1}$ so that it is an inverse of the linearly-independent (sub-)space of ${\cal N}$.
We then consider the potential given by
\begin{eqnarray}
U_{2N}(\vec{r},\vec{r}') = 
\sum_{W_i,W_j < W_{\rm th}} 
(E^{W_i}_{2N}-H_0) \phi_{2N}^{W_i}(\vec{r}) {\cal N}^{-1}_{W_i,W_j} \overline{\phi_{2N}^{W_j}(\vec{r}')} .
\label{eq:U_e-indep}
\end{eqnarray}
It is evident that this (trivially energy-independent) 
potential satisfies the Schr\"odinger equation (\ref{eq:Sch_2N:tindep})
for all $W < W_{\rm th}$.
It is also possible to show that one can construct energy-independent potential 
even above inelastic threshold~\cite{Aoki:E-indep:inelastic}.
This ``existence proof'' of energy-independent potential 
plays an essential role in the HAL QCD method.

Several remarks are in order.
First, a potential itself is not an observable
and is not unique.
It depends on the definition of NBS wave functions,
e.g., the choice of the nucleon operator $N$.
One can also consider another form of $U_{2N}$ instead of Eq.~(\ref{eq:U_e-indep}).
Recall, however, that 
physical observables calculated from different potentials, 
such as phase shifts, 
are unique by construction. 
Therefore, while there exists a ``scheme'' dependence in a potential,
it is not a problematic ambiguity.
Rather, it is a freedom at our disposal,
analog to the freedom to choose a ``scheme'' in perturbative calculations.
Recall also that modern nuclear calculations
often take advantage of the freedom to define the potential~\cite{Bogner:2003wn,Bogner:2006pc}.
Second, in practical lattice calculations,
it is difficult to handle the non-locality of the potential directly,
since Eq.~(\ref{eq:U_e-indep}) requires NBS wave functions 
at all energies below $W_{\rm th}$.
To proceed, we employ the derivative expansion of the potential~\cite{okubo-marshak},
%
$
U_{2N}(\vec{r},\vec{r}') =
\left[ V_C(r) + V_T(r) S_{12} + V_{LS}(r) \vec{L}\cdot \vec{S} + {\cal O}(\nabla^2) \right]
\delta(\vec{r}-\vec{r}') , 
$
%
%
where $V_C$, $V_T$ and $V_{LS}$ are the central, tensor and spin-orbit potentials, respectively,
with the tensor operator $S_{12}$.
In this way, we can determine the potentials order by order with 
a realistic number of NBS wave functions determined on the lattice.
In Ref.~\cite{Murano:2011nz}, the convergence of the derivative expansion is examined 
in parity-even channel,
and it is shown that the leading terms, $V_C$ and $V_T$, dominate the potential at low energies.

Finally, let us address several advantages of the approach based on the potential.
The first point is that it is a convenient framework to understand the physics.
One of the examples is given in Sec.~\ref{sec:YN}, where we discuss the origin of 
repulsive core from the viewpoint of Pauli exclusion principle.
A potential is also a useful tool to study many-body systems,
since various many-body techniques have been developed in nuclear physics based on potentials.
This could serve as an alternative approach to dense systems,
for which direct lattice QCD simulations are difficult due to the 
sign problem.
Another advantage is that a potential is a localized object, and thus
the finite volume artifact is better under control.
Actually, once a potential is obtained in a finite lattice box,
we can solve the Schr\"odinger equation in infinite volume.
Last but not least, 
it is possible to extend the HAL QCD method
so that the ground state saturation is not required.
This point is a significant advantage over the traditional L\"uscher's method,
and will be more elaborated in the next section.

\vspace*{-1mm}
\section{Challenges toward lattice simulations with realistic setup}
\label{sec:challenges}

\vspace*{-2mm}

\subsection{The signal to noise issue}
\label{subsec:S/N}

In order to study nuclear physics from lattice QCD,
we ultimately have to perform simulations at the physical point
with sufficiently large volumes.
Toward this direction, a major challenge that lies ahead 
is the so-called S/N issue.
This issue arises since lattice simulations usually rely on the ground state saturation,
which is in principle achieved by 
taking an infinitely large Euclidean time separation 
in the correlation function of concern.
For instance, the NBS wave function of the ground state of the 2N system, $\phi_{2N}^{W_0}(\vec{r})$,
is extracted from the four-point correlator as
\begin{eqnarray}
\label{eq:4pt_2N}
G_{2N} (\vec{r},t-t_0)
&\equiv&
\frac{1}{L^3}
\sum_{\vec{R}}
\langle 0 |
          (N(\vec{R}+\vec{r}) N (\vec{R}))(t)\
\overline{(N' N')}(t_0)
| 0 \rangle \ \
\xrightarrow[t \gg t_0]{} \ \ 
A_{2N}^{W_0} \phi_{2N}^{W_0} e^{-W_0(t-t_0)} , \\
\quad
\phi_{2N}^{W_0}(\vec{r}) &=& \langle 0 | N(\vec{r}) N(\vec{0}) | 2N, W_0\rangle_{\rm in} ,
\quad
A_{2N}^{W_0} = _{\rm in}\!\langle 2N, W_0 | \overline{(N' N')} | 0 \rangle , 
%
%
\end{eqnarray}
where 
$W_0$ denotes the energy of the ground state,
$N$ ($N'$) the nucleon operator in the sink (source).
In the practical lattice calculation, however, it is notoriously difficult
to achieve the ground state saturation
for multi-baryon systems. 
In fact, for the correlation function of $A$-nucleon systems,
the S/N becomes~\cite{Lepage:1989hd} 
%
%
$
S/N \sim \exp\left[ -A (m_N - 3 m_\pi/2 ) (t-t_0) \right]
$ 
for 
$t \gg t_0$, 
%
%
where $m_\pi$ is the pion mass.
To make matters worse,
there exists another problem for multi-baryon systems on the lattice, i.e.,
the energy splitting between the ground state and excited (scattering) states
becomes (too) smaller for larger lattice volume.
For instance, the minimum splitting of the 2N system 
in a lattice box with a spacial size of $L$ is
%
%
$
\Delta E \simeq \vec{p}^2_{\rm min} / m_N = (2\pi)^2 / (m_N L^2).
$
%
%
If we want to carry out simulations with $L\sim 10$ fm at the physical point,
$\Delta E \sim 15$ MeV and thus $t/a \gg 100$ 
may be required with a lattice spacing of $a \sim 0.1$ fm.
While one may employ techniques such as diagonalization of correlation function matrix
to ameliorate this problem,
this remains a serious issue as the volume gets larger.

In Ref.~\cite{HALQCD:2012aa}, a novel approach to resolve this issue is proposed,
by extending the HAL QCD method.
The essential point is that 
the (elastic) scattering states with different energies
on the lattice are governed by the same potential,
since an ``energy-independent'' (non-local) potential is utilized in the HAL QCD method.
With this realization, one can construct the method 
in which an potential is extracted
without relying on the ground state saturation.
More specifically, we introduce an imaginary-time NBS wave function defined by
%
%
$
\psi_{2N}(\vec{r},t) \equiv G_{2N} (\vec{r},t) / e^{-2m_N t} ,
$
%
%
and consider the time-dependent Schr\"odinger equation,
\begin{eqnarray}
%
H_0 \psi_{2N}(\vec{r},t)
+ \int d\vec{r}' U_{2N}(\vec{r},\vec{r}') \psi_{2N}(\vec{r}',t)
= \left( 
- \frac{\partial}{\partial t} 
+ \frac{1}{4m_N} \frac{\partial^2}{\partial t^2} 
\right)
\psi_{2N}(\vec{r},t) .
\label{eq:Sch_2N:tdep}
\end{eqnarray}
It is easy to see that Eq.~(\ref{eq:Sch_2N:tdep}) is consistent with
Eq.~(\ref{eq:Sch_2N:tindep}),
even when there exist contributions from excited states in 
$\psi_{2N}(\vec{r},t)$.
Therefore, in this ``time-dependent HAL QCD method'',
the ground state saturation is not required any more
as far as contaminations from states above inelastic threshold are suppressed.
(For extension of the time-dependent HAL QCD method above inelastic threshold, 
 see Ref.~\cite{Aoki:E-indep:inelastic}).
The effectiveness of this new method
have been examined in explicit numerical simulations as well.
For 2N systems~\cite{HALQCD:2012aa} and $I=2$ $\pi\pi$ system~\cite{Kurth:lat2012},
it is confirmed that reliable potentials and phase shifts can be extracted
even with the presence of excite state contributions.


\subsection{The computational cost issue}
\label{subsec:cost}

Another major challenge in multi-baryon systems on the lattice
is the computational cost of the correlation functions.
In particular, it is well known that the cost of the contraction
is exceptionally enormous for larger baryon number $A$,
since 
(i) the number of quark permutations (Wick contractions) grows  
factorially with $A$
and
(ii) the contraction of color/spinor degrees of freedom (DoF) becomes 
exponentially large for large $A$.
While there has been significant progress toward reducing this computational 
cost~\cite{Yamazaki:2009ua, Ishii:2006ec, Aoki:2009ji, Doi:2010yh, Doi:2011gq, Kaplan:DWF10yrsTalk},
it continues to remain the most time-consuming part of the calculation,
particularly for $A > 2$.%

On this issue, we recently developed a novel algorithm,
called ``unified contraction algorithm,''
which achieves a drastic reduction of the computational cost~\cite{Doi:2012xd}.
Essential idea is
to consider the Wick contractions
and the color/spinor contractions simultaneously.
In fact, 
if the quarks of the same flavor have 
the same space-time smearing function 
at the sink and/or source,
a permutation of quark operators is equivalent to 
a permutation of color and spinor indices of the corresponding quark.
Since color/spinor indices are dummy indices in the color/spin contractions,
we can carry out full permutations (i.e., Wick contractions) 
in advance of the lattice simulation.
This procedure amounts to 
preparing a unified index list for Wick and color/spinor contractions.
If there exists any redundancy and/or cancellation among contributions 
in the original contraction,
they are automatically consolidated when constructing the unified index list,
thus significant speedup is achieved.

The method is rather general and may be applied 
to, e.g., the straightforward algorithm,
its extension with the determinant algorithm~\cite{Kaplan:DWF10yrsTalk}
and the block algorithm (see Ref.~\cite{Doi:2012xd} for these definitions).
The last one, in which the baryon block is constructed at this sink,
is particularly useful to achieve a large overlap with the state of interest
and also to calculate the NBS wave functions.
Explicit study~\cite{Doi:2012xd} shows that 
a significant reduction in the computational cost is achieved, e.g., 
by a factor of 192 for $^3$H and $^3$He nuclei, 
a factor of 20736 for the $^4$He nucleus
and a factor of ${\cal O}(10^{11})$ for the $^8$Be nucleus
without assuming isospin symmetry.
A further reduction is possible by exploiting isospin symmetry, 
and/or interchange symmetries associated with sink baryons, if such symmetries exist.
This algorithm is also useful to study YN, YY forces,
where the calculation of coupled channel correlators
require considerable computational resources.

\vspace*{-2mm}
\section{Two-nucleon forces}
\label{sec:2NF}

\vspace*{-2mm}

\subsection{Central and tensor forces in parity-even channel}
\label{subsec:2NF:Peven}

\begin{figure}[t]
\vspace*{-5mm}
\begin{minipage}{0.48\textwidth}
\begin{center}
\includegraphics[angle=-90,width=0.85\textwidth]{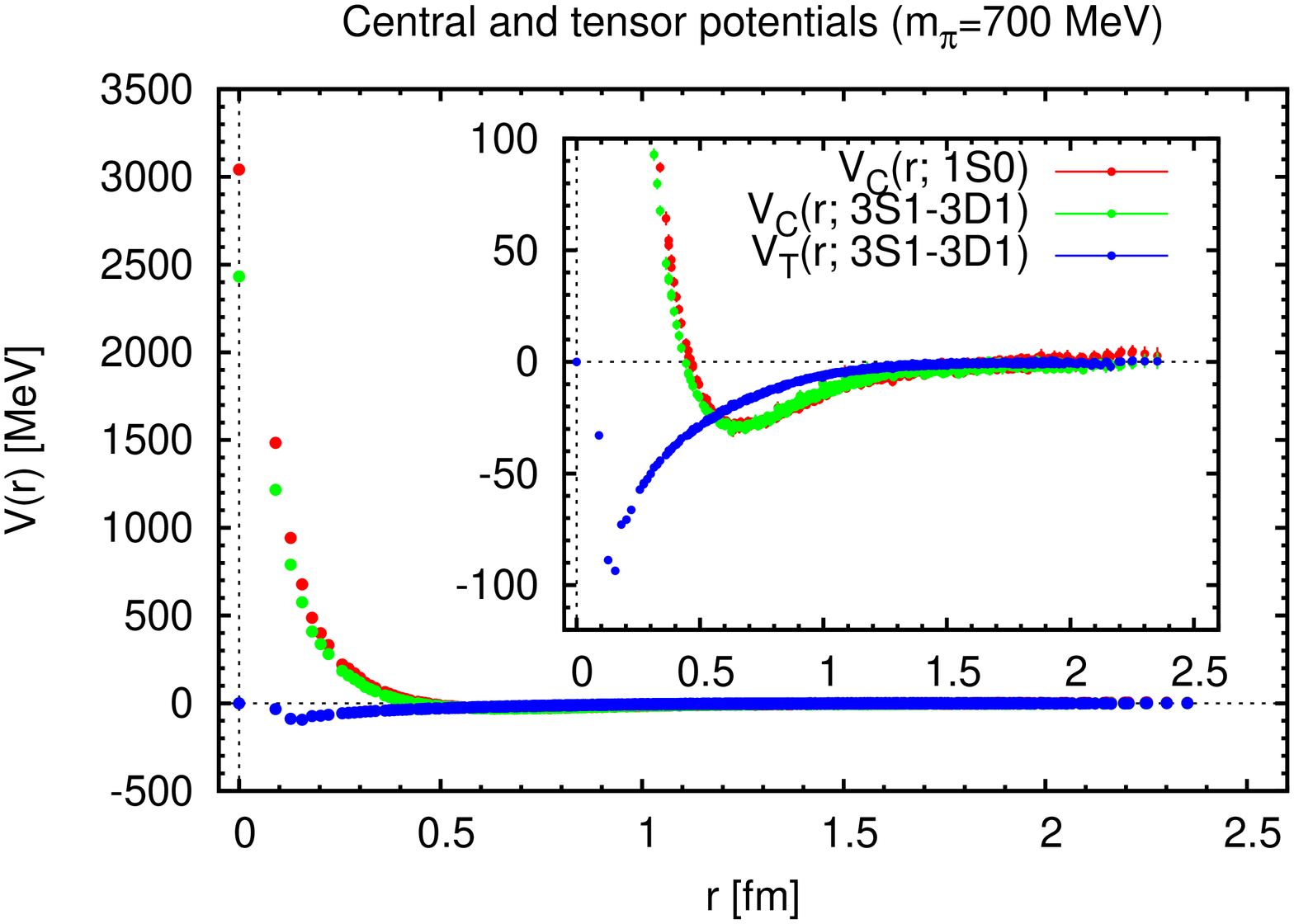}
\caption{
\label{fig:2NF:Peven}
Parity-even 2NF 
in $^1S_0$ and $^3S_1-^3D_1$ channels
obtained on the lattice at $m_\pi = 0.70$ GeV.
}
\end{center}
\end{minipage}
\hfill
\begin{minipage}{0.48\textwidth}
\begin{center}
\includegraphics[angle=-90,width=0.85\textwidth]{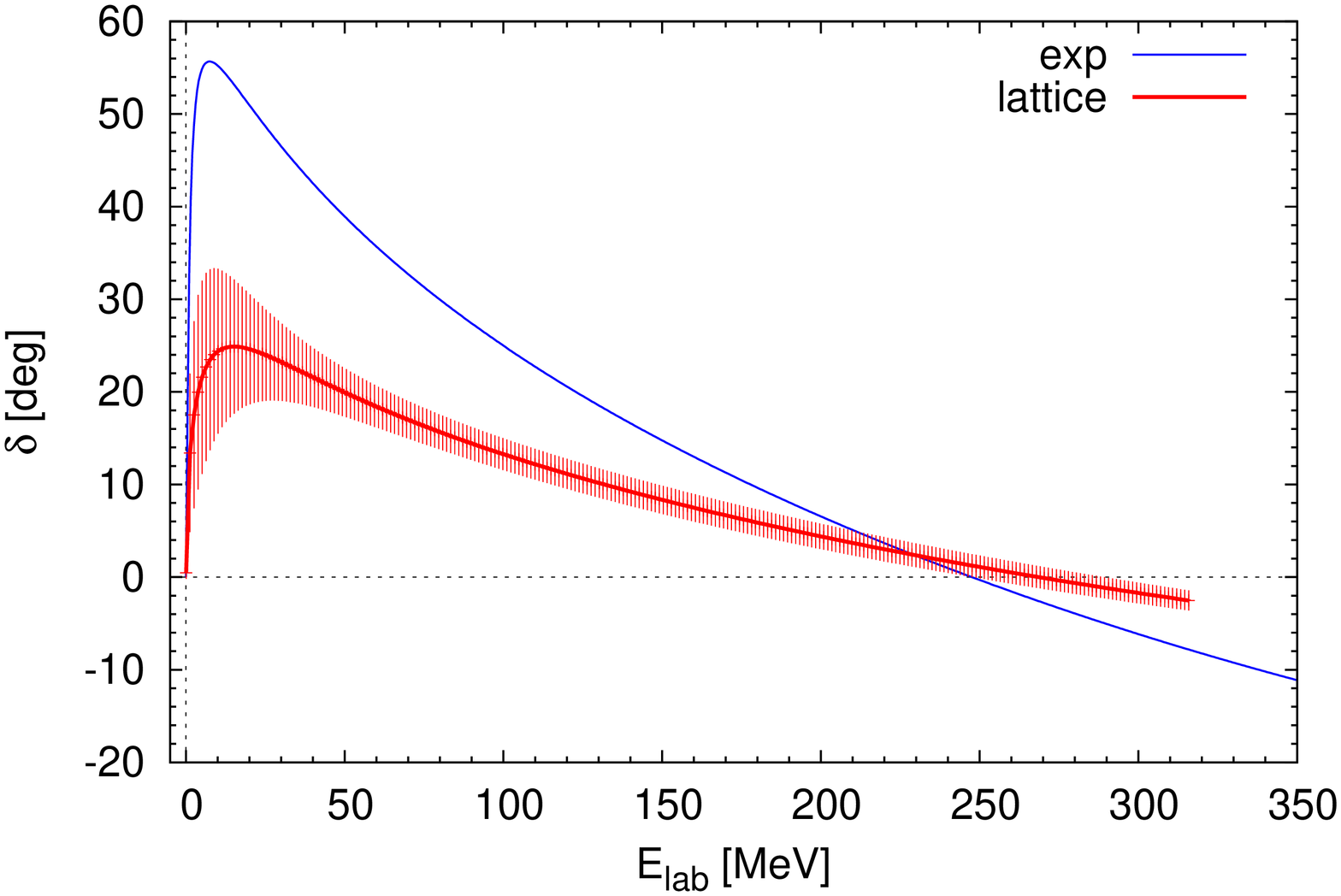}
\caption{
\label{fig:NN:phase}
The obtained phase shift in $^1S_0$ channel
in the laboratory frame, 
with experimental data~\cite{HALQCD:2012aa}.
}
\end{center}
\end{minipage}
\end{figure}

We present the results for parity-even 2NF in $^1S_0$ and $^3S_1-^3D_1$ channels,
which corresponds to ``dineutron'' and ``deuteron'' channels, respectively.
We determine the potentials up to the fist order in the derivative expansion, 
namely, a central force in $^1S_0$ channel
and central and tensor forces in $^3S_1-^3D_1$ channel.
%
%
Quenched QCD~\cite{Ishii:2006ec,Aoki:2009ji} and 
full QCD~\cite{HALQCD:2012aa,Aoki:2012tk} studies have been performed,
and the results from the latter are presented in this report.
We employ $N_f = 2+1$ 
clover fermion configurations
generated by PACS-CS Collaboration~\cite{Aoki:2008sm}.
The lattice spacing is $a \simeq 0.091$ fm 
and 
the lattice size of $V = L^3 \times T = 32^3\times 64$
corresponds to
(2.9 fm)$^3$ box in physical spacial size.
For quark masses, 
we take three hopping parameters at the unitary point
as
$\kappa_{ud} = 0.13700, 0.13727, 0.13754$ for $u$, $d$ quark masses, 
and $\kappa_s = 0.13640$ for $s$ quark mass.
The hadron masses at each $\kappa_{ud}$  correspond to
$m_\pi \simeq 701, 570, 411$ MeV 
and $m_N \simeq 1584, 1412, 1215$ MeV, respectively.
For a nucleon operator, $N$,
we employ the standard operator,
$
N(x) = \epsilon_{abc} (q_a^T(x) C \gamma_5 q_b(x)) q_c(x)
$,
at both sink and source,
and wall quark source is used with Coulomb gauge fixing.

In Fig.~\ref{fig:2NF:Peven},
we plot the results for the central potential in $^1S_0$ channel
and the central and tensor potentials in $^3S_1-^3D_1$ channel
obtained at $m_\pi = 0.70$ GeV.
In the latter, the central and tensor potentials
are extracted from the coupled channel Schr\"odinger equation
between S-wave and D-wave components of the NBS wave function.
The obtained potentials reproduce the qualitative features
of the phenomenological potentials,
namely, attractive wells at long and medium
distances, central repulsive cores at short distance
and negative tensor force.

To obtain the results for physical observables,
we fit the potential with a multi-Gaussian function,
and solve the Schr\"odinger equation in infinite volume.
We find that both of dineutron and deuteron are not bound.
In Fig.~\ref{fig:NN:phase}, we show the results for the scattering phase shift
in $^1S_0$ channel at $m_\pi = 0.70$ GeV.
A qualitative feature of the experimental data is well reproduced,
though the strength is weaker, most likely due to the heavy pion mass.
The scattering length obtained from the derivative of the
phase shift at $k=0$ becomes 
$a(^1S_0) = \lim_{k\rightarrow 0}\tan \delta(k)/k = 1.6(1.1)$ fm,
which is compared to the experimental value
$a^{\rm exp}(^1S_0) \simeq 20$ fm.

\subsection{Central, tensor and spin-orbit forces in parity-odd channel}
\label{subsec:2NF:Podd}

\begin{figure}[t]
\vspace*{-3mm}
\begin{center}
\includegraphics[width=0.32\textwidth]{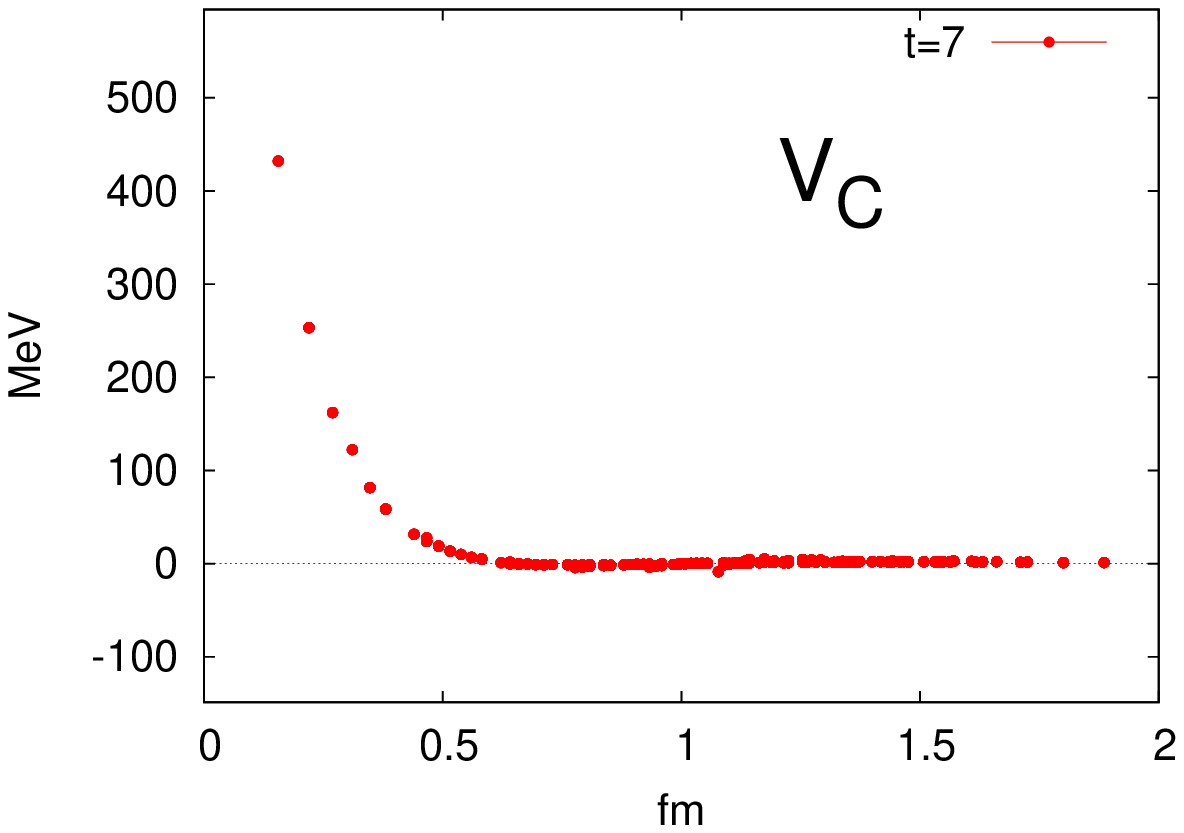}
\includegraphics[width=0.32\textwidth]{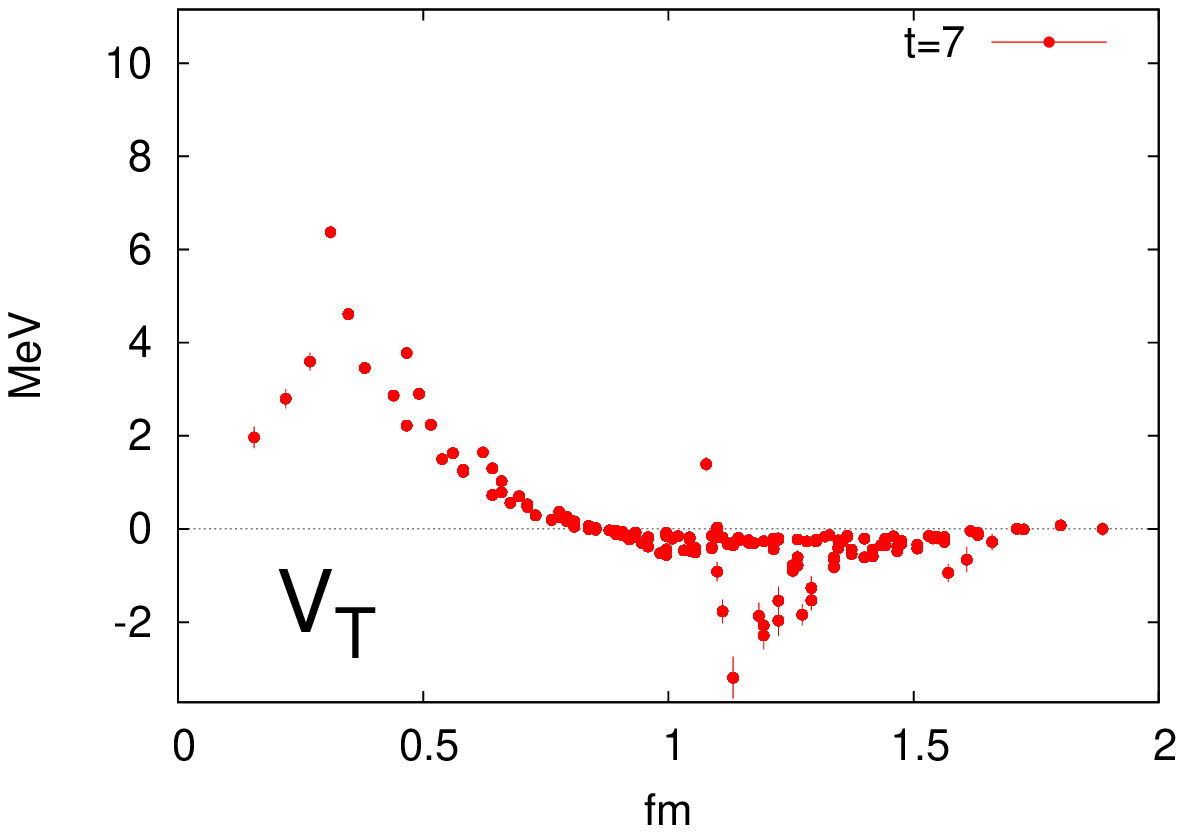}
\includegraphics[width=0.32\textwidth]{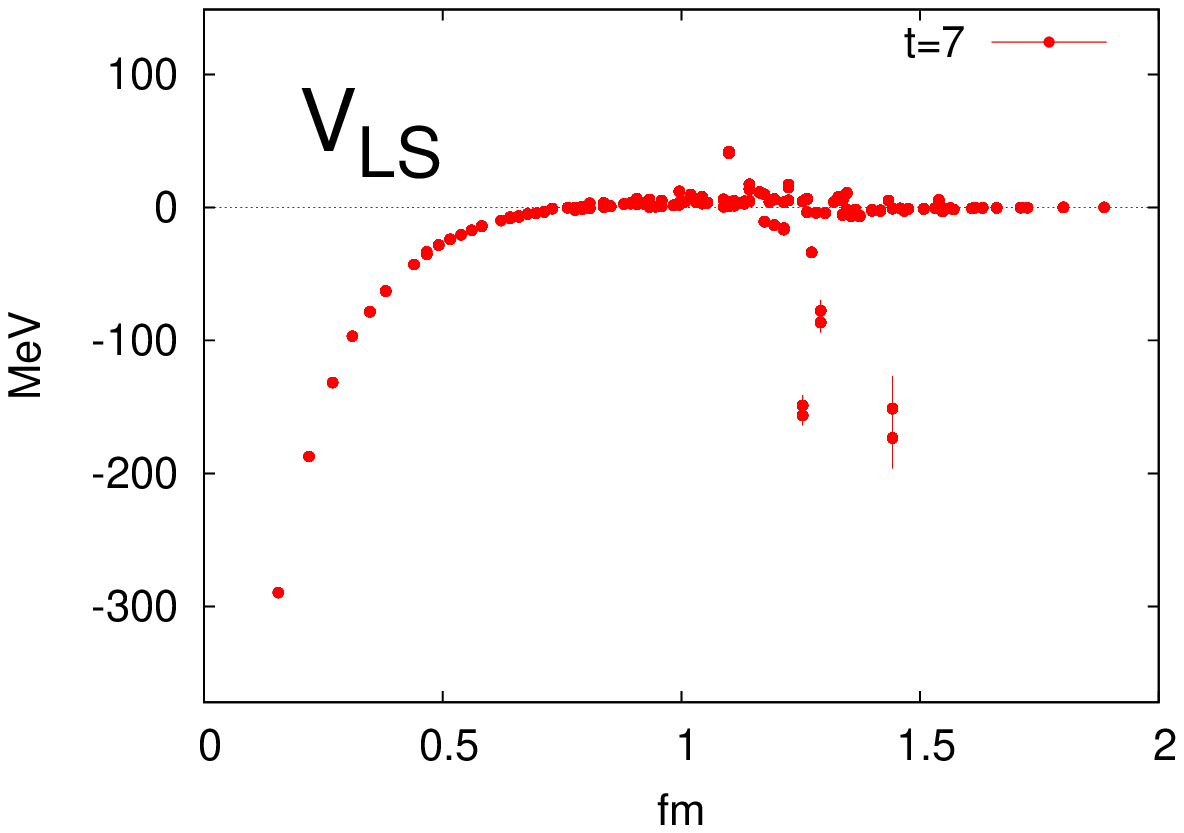}
\caption{
\label{fig:2N:Podd}
Parity-odd 2NF in $^3P_0$, $^3P_1$ and $^3P_2-^3F_2$ channels
obtained on the lattice at $m_\pi=1.1$ GeV.
Left, middle, and right figures show central, tensor and spin-orbit forces, 
respectively~\cite{Aoki:2012tk}. 
}
\end{center}
\vspace*{-2mm}
\end{figure}

Nuclear forces in parity-odd channels 
are important 
not only in P-wave 2N scattering,
but also in many-nucleon systems.
In addition,
spin-orbit forces
attract a great deal of interest recently,
since those in $^3P_2$ channel
are considered to be relevant in superfluidity in neutron stars.
They are also partly responsible for the magic numbers of nuclei
in the nuclear shell model.

The lattice QCD study for parity-odd 2NF is more involved than
the study of parity-even 2NF,
since non-zero relative momentum has to be injected in the system.
We employ a nucleon source operator with a momentum,
$
N' = \sum_{\vec{x}_1,\vec{x}_2,\vec{x}_3}
\epsilon_{abc} (q_a^T(\vec{x}_1) C \gamma_5 q_b(\vec{x}_2)) q_c(\vec{x}_3) f(\vec{x}_3)
$
with $f(\vec{x}) = \exp [\pm 2\pi i x_k / L]$, $k=1,2,3$.
%
We consider NBS wave functions in $J^P = A_1^-, T_1^-, T_2^-$ channels
in the cubic group,
which correspond to 
$^3P_0$, $^3P_1$, $^3P_2-^3F_2$ in the continuum limit.

Numerical calculations  are performed  by 
employing
$N_f=2$ dynamical 
configurations
with mean field improved clover fermion 
and 
renormalization-group improved
gauge action
generated by CP-PACS Collaboration~\cite{Ali_Khan:2001tx}.
The lattice spacing is $a^{-1} = 1.269(14)$ GeV
and the lattice size of $V = L^3 \times T = 16^3\times 32$
corresponds to
(2.5 fm)$^3$ box in physical spacial size.
For $u$, $d$ quark masses, 
we take the hopping parameter at the unitary point
as
$\kappa_{ud} = 0.13750$,
which corresponds to
$m_\pi = 1.1$ GeV and
$m_N = 2.2$ GeV.

We determine the central ($V_C$), tensor ($V_T$) and spin-orbit potentials ($V_{LS}$)
from the (coupled) Schr\"odinger equations in $J^P = A_1^-, T_1^-, T_2^-$ channels.
Shown in Fig.~\ref{fig:2N:Podd} are the preliminary results 
for these potentials from lattice QCD.
Their features qualitatively agree with those in phenomenological potentials,
as
(i)  $V_C$  has
repulsive core at  short distance,
(ii) $V_T$ is  positive and very small,
and
(iii)  $V_{LS}$  is large  and  negative  at  short distance.

\section{Hyperon forces}
\label{sec:YN}

\begin{figure}[t]
\vspace*{-5mm}
\begin{center} 
 \includegraphics[width=0.32\textwidth]{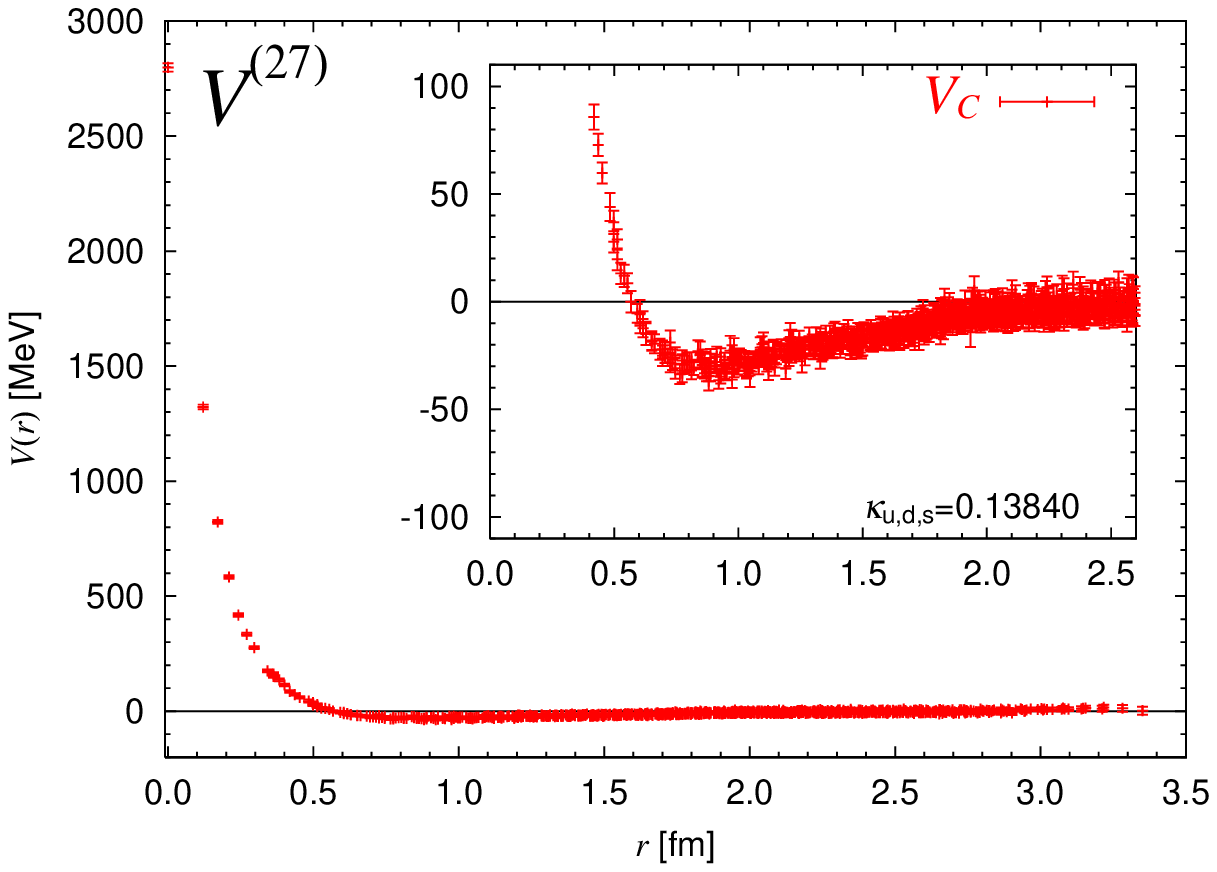}
 \includegraphics[width=0.32\textwidth]{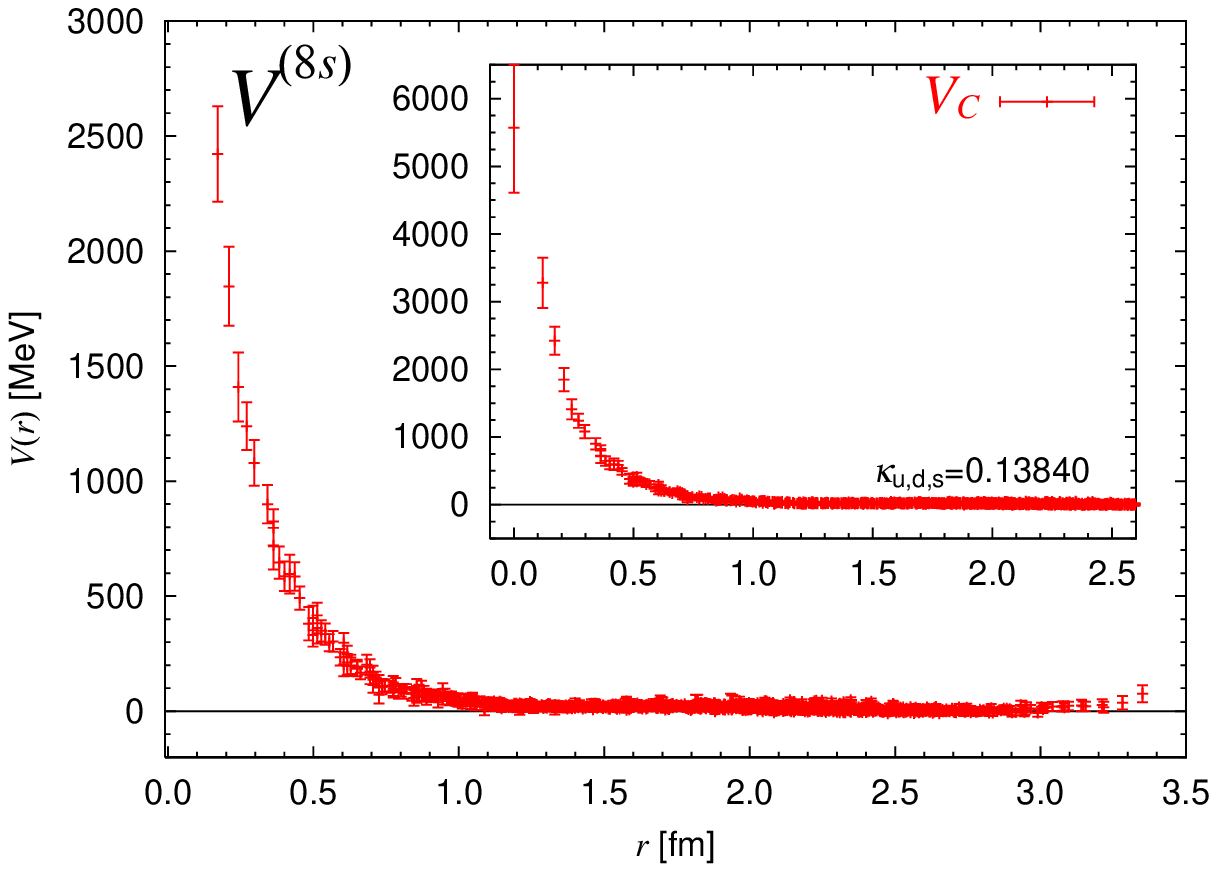}
 \includegraphics[width=0.32\textwidth]{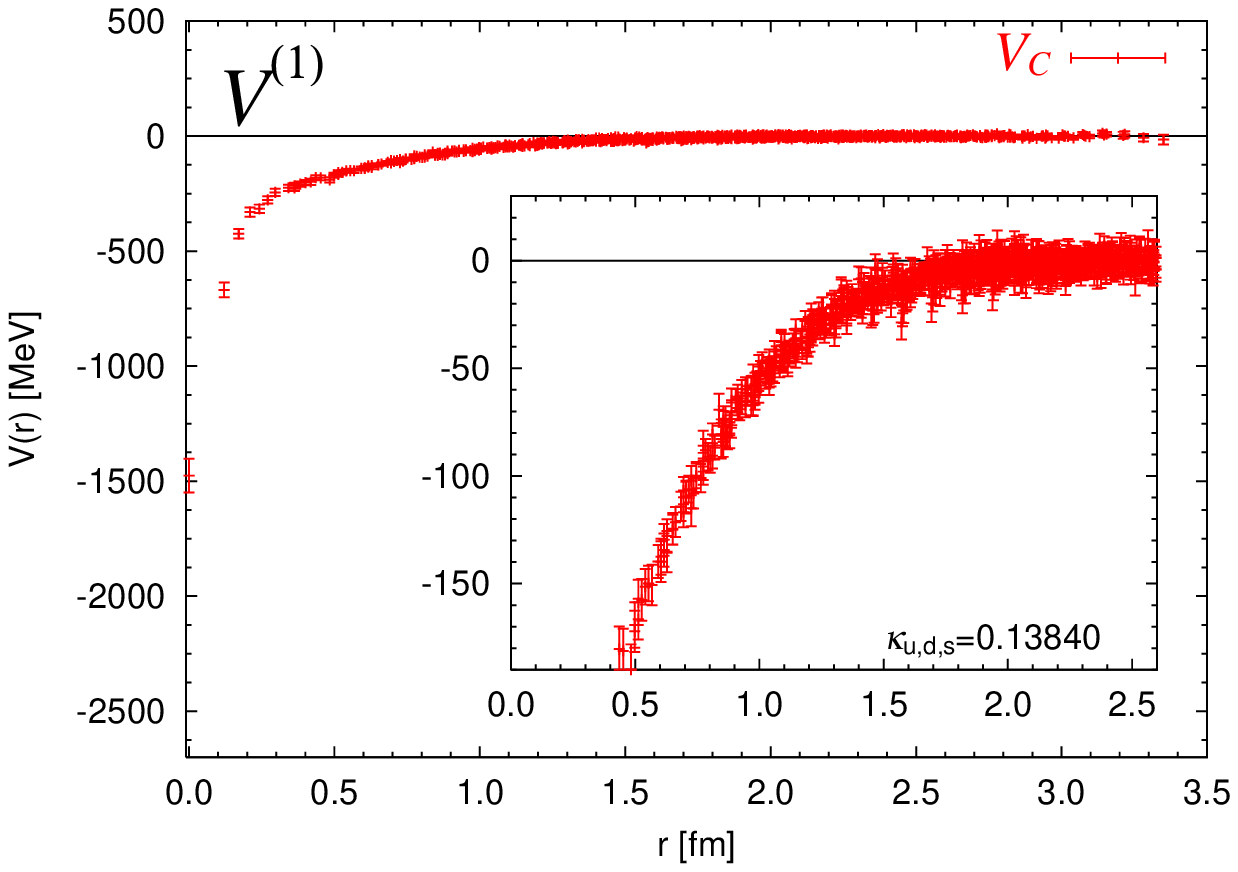}\hfill
 \includegraphics[width=0.32\textwidth]{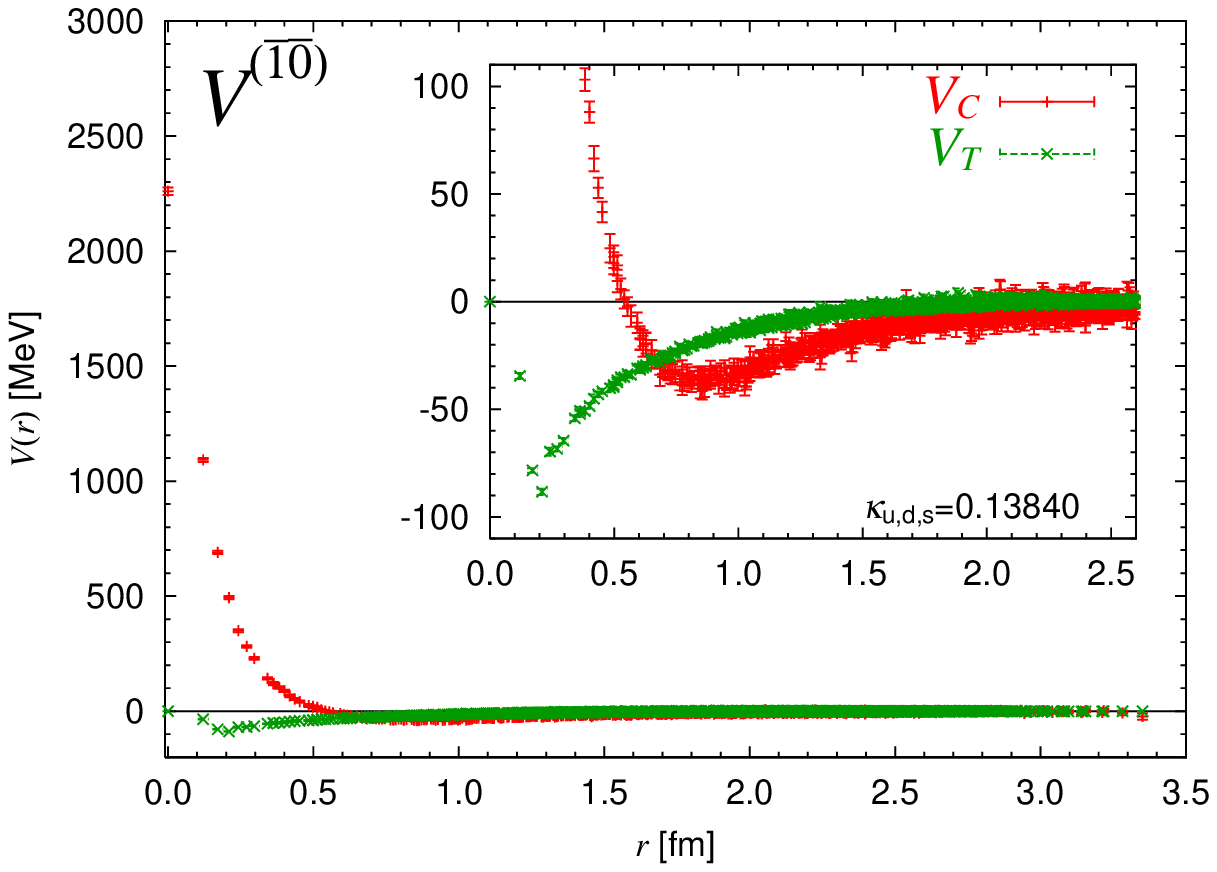}
 \includegraphics[width=0.32\textwidth]{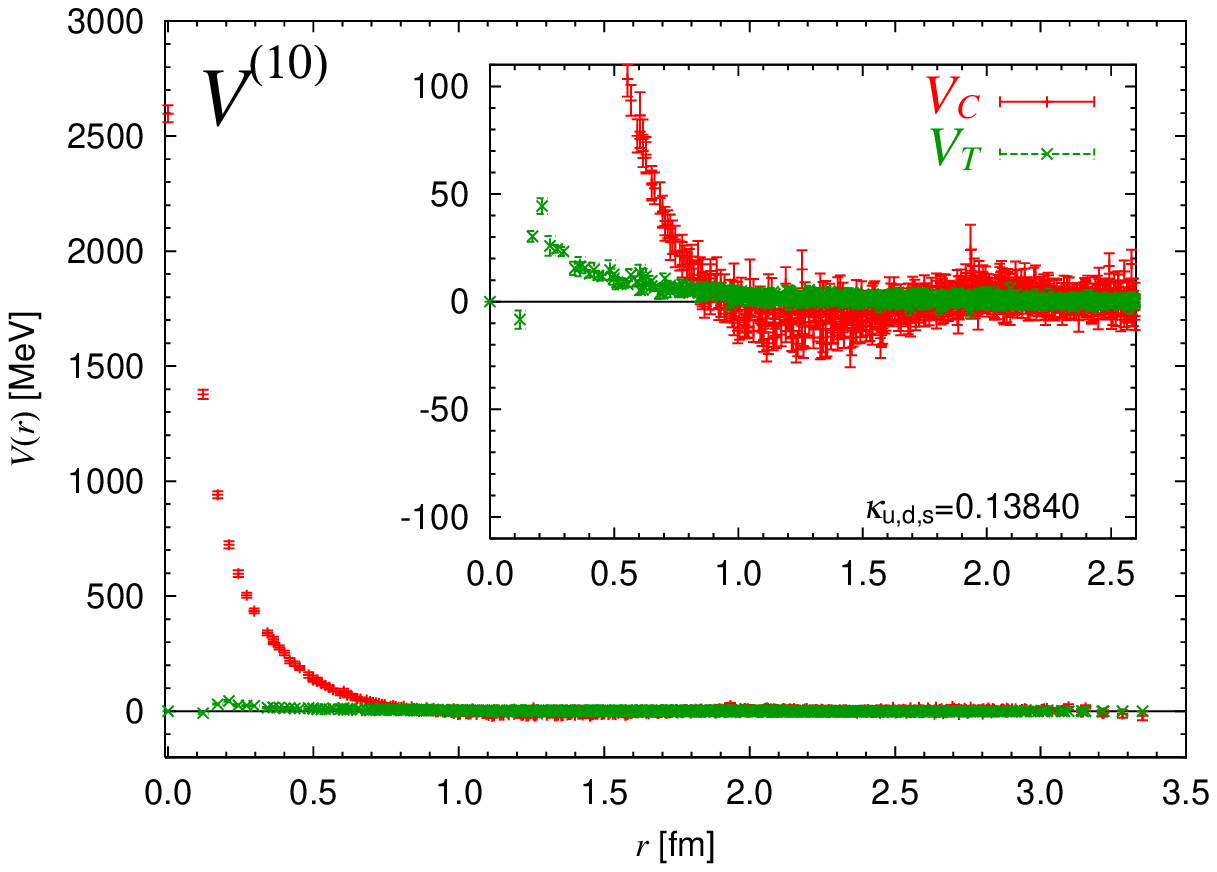}
 \includegraphics[width=0.32\textwidth]{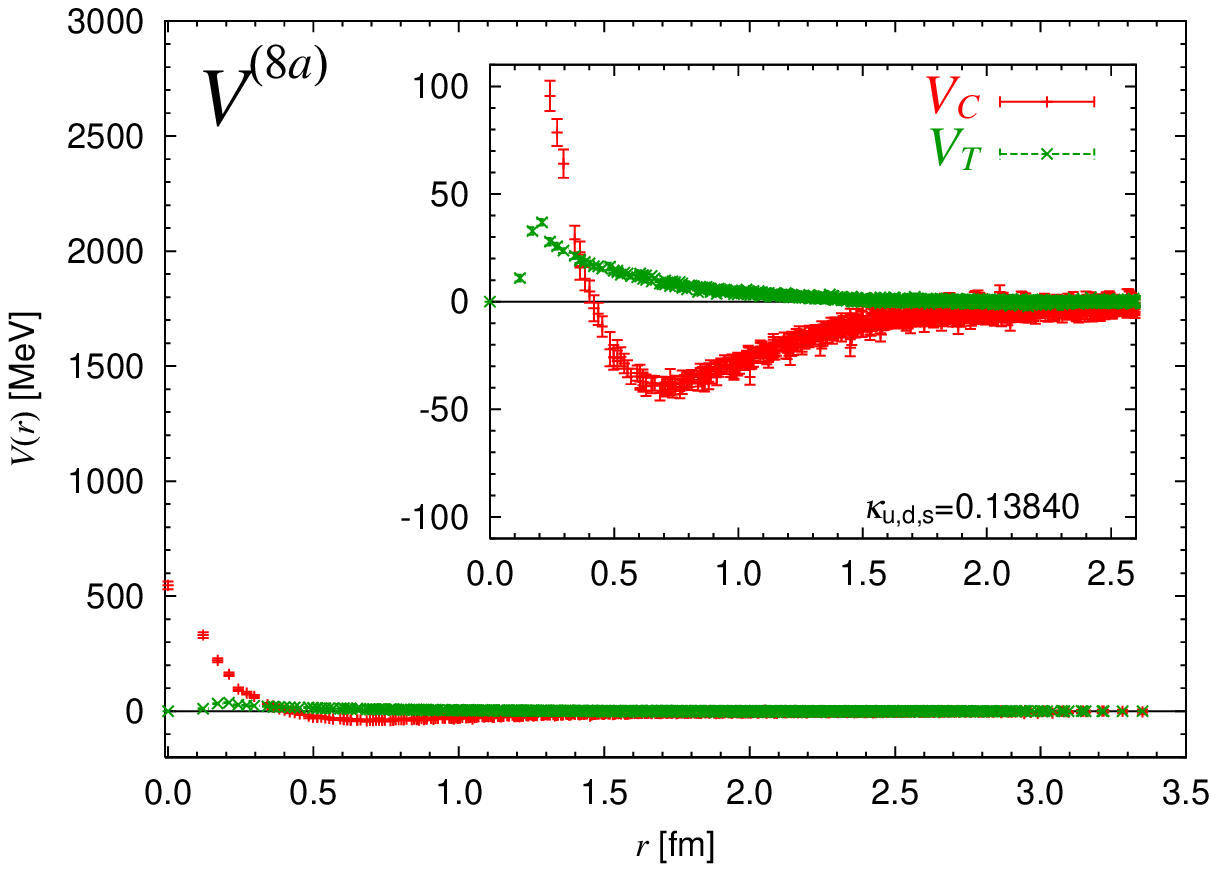}
\caption{
The 2B potentials in 
${\bf 27}$  (upper left),
${\bf 8_s}$ (upper middle),
${\bf 1}$   (upper right),
$\overline{\bf 10}$  (lower left),
${\bf 10}$  (lower middle) and
${\bf 8_a}$ (lower right) flavor representation,
obtained at 
$M_{\rm ps}=469$ MeV on the lattice~\cite{Inoue:2011ai}.
\vspace*{-5mm}
 } 
\label{fig:su3limitB}
\end{center}
\end{figure}

Hyperon is a new DoF in nuclear physics
and their interactions 
are crucial to understand hypernuclei
and the structure of neutron star cores.
They are also essential inputs to explore exotic multi-quark states such as the $H$-dibaryon.
While YN/YY scattering experiments are difficult because of the short life-time of hyperons,
lattice QCD is suitable framework to determine hyperon interactions.
The first calculations for hyperon potentials are performed
in Ref.~\cite{Nemura:2008sp} for $p\Xi^0$ system in quenched simulations.
In this report, we present the latest results for generalized baryon forces
in $N_f = 3$ full QCD simulations~\cite{Inoue:2010hs,Inoue:2010es,Inoue:2011ai}, 
and the results in $N_f = 2 + 1$ full QCD simulations~\cite{Sasaki:2010bi}.

\subsection{Generalized baryon forces in the flavor SU(3) limit and bound $H$-dibaryon}
\label{subsec:2BF:su3}

In order to grab the insight of physics, 
it is convenient to consider the generalized baryon forces (NN, YN, YY forces)
in the flavor SU(3) limit. 
In this limit, two-baryon (2B) systems composed of spin $1/2$ flavor-octet baryon
can be classified by irreducible representation of SU(3) as
\begin{eqnarray}
{\bf 8}\otimes {\bf 8} = \underbrace{{\bf 27}\oplus {\bf 8_s}\oplus {\bf 1}}_{\rm symmetric}\oplus 
 \underbrace{\overline{\bf 10}\oplus {\bf 10}\oplus {\bf 8_a}}_{\rm anti-symmetric},
\end{eqnarray}
where "symmetric" and "anti-symmetric" denotes the symmetry under the exchange of two baryons.
For the system with 
S-wave, Pauli principle 
imposes {\bf 27}, ${\bf 8_s}$ and {\bf 1}
to be spin-singlet ($^1S_0$), while  $\overline{\bf 10}$, {\bf 10}  and ${\bf 8_a}$ to be spin-triplet ($^3S_1-^3D_1$).
We note that, 2NF corresponds to either ${\bf 27}$ or $\overline{\bf 10}$,
and other 4 representations are purely unique interactions with the presence of hyperons.

We generate $N_f=3$ dynamical configurations
with ${\cal O}(a)$ improved clover fermion action
and 
renormalization-group improved gauge action
on a $32^3\times 32$ lattice at $a\simeq 0.12$ fm,
and at five values of quark hopping parameters,
which corresponds to 
$(M_{\rm ps}, M_B) =$
(1170.9(7),~2274(2)), 
(1015(1),~2030(2)),
(837(1),~1748(1)), 
(673(1),~1485(2)) and 
(468.6(7),~1161(2)) in unit of MeV,
where $M_{\rm ps}$ and $M_B$ denote 
the masses of the octet pseudoscalar (PS) meson and the octet baryon, respectively.
In the calculation of the NBS wave functions,
we construct a 2B operator with appropriate Clebsch-Gordan coefficients
to respect the irreducible representation of interest.

In Fig.~\ref{fig:su3limitB},
we show the obtained potentials 
at  $M_{\rm ps} = 469$ MeV
for each flavor representation~\cite{Inoue:2011ai}.
The upper panels show central forces in the spin-singlet channel, 
while the lower panels give central and tensor forces in the spin-triplet channel.
%
%
%
What is noteworthy is that 
potentials are highly dependent on the flavor representation.
In particular, compared to 2NF sectors, $V^{\bf (27)}_C$ and $V^{\bf (\overline{10})}_{C,T}$,
$V^{\bf (10)}_C$ has a stronger repulsive core
and a weaker attractive pocket.
Furthermore, 
$V^{({\bf 8}_s)}_C$ has a very strong repulsive core
among all 6 channels, while $V^{({\bf 8}_a)}_C$ has a very weak repulsive core. 
In contrast to all other cases,  $V^{\bf (1)}_C$ has attraction at short distances instead of repulsion.
These features are found to be well explained 
from the viewpoint of the Pauli exclusion principle in the quark level~\cite{Oka:2000wj}. 
Such agreements between the lattice data and the quark model suggest
that the quark Pauli exclusion plays an essential role for the repulsive core in 2B systems.

The potential in {\bf 1} channel is particularly interesting,
since the existence of an exotic $H$-dibaryon was proposed in this channel,
and lattice QCD results show the existence of an attractive core.
We fit the flavor singlet potential
and solve the Schr\"odinger equation in infinite volume.
It turns out that, at each quark mass, there is only one bound state with binding energy of 20--50 MeV,
with smaller binding energy for smaller quark mass.
On the other hand, there appears no bound state 
in the 27-plet channel  (``dineutron'')
or  the $\overline{10}$-plet channel ("deuteron" ) in the present range of quark masses.
We note that, since the binding energy of $H$-dibaryon
is comparable to the splitting between physical hyperon masses
and not so sensitive to quark mass, 
there may be a possibility of weakly bound or resonant $H$-dibaryon even in the real world. 
To make a definite conclusion, however, 
the $\Lambda\Lambda-N\Xi-\Sigma\Sigma$ coupled channel analysis is necessary for 
$H$-dibaryon
in the $N_f = 2 + 1$ lattice QCD simulations, as will be discussed in the next section.

\vspace*{-2mm}
\subsection{Coupled channel hyperon forces with flavor SU(3) breaking}
\label{subsec:2BF:coupled}

\begin{figure}[t]
\vspace*{-6mm}
\begin{center}
  \includegraphics[width=0.35\textwidth]{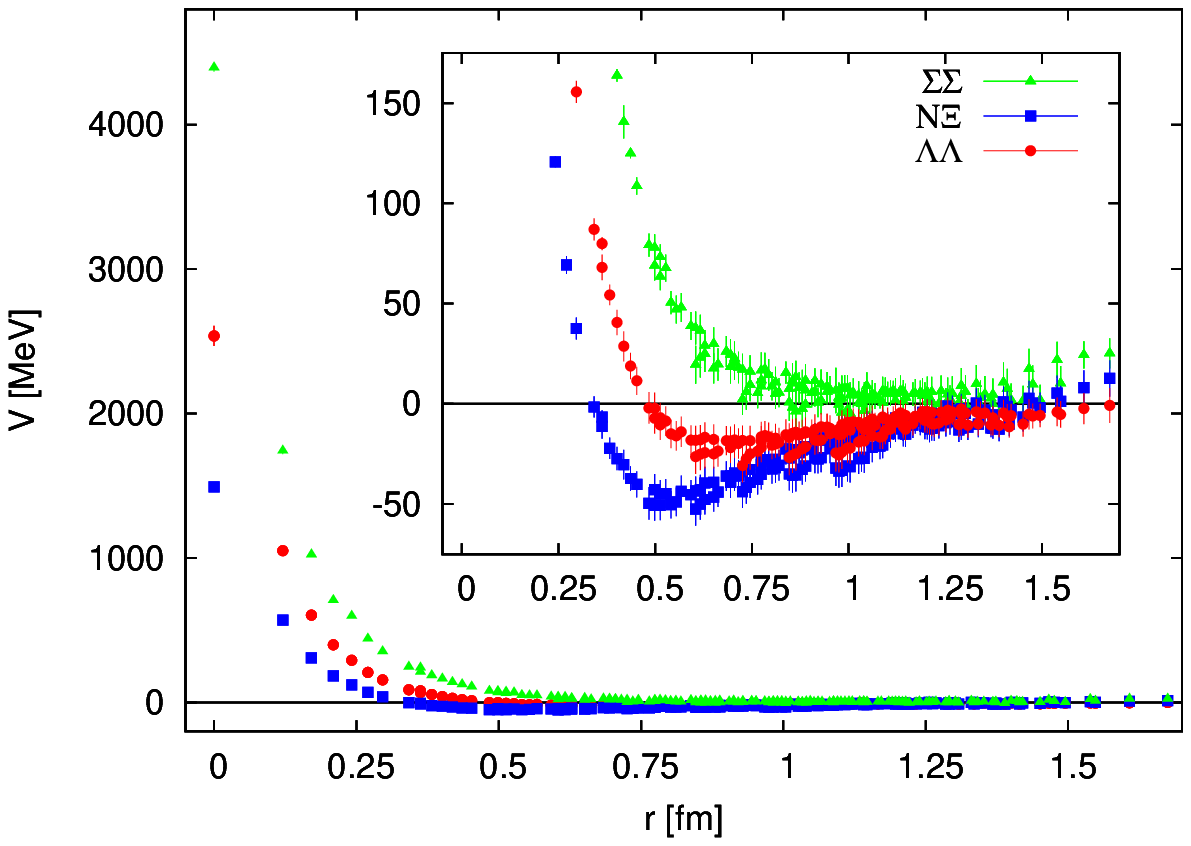}
  \includegraphics[width=0.35\textwidth]{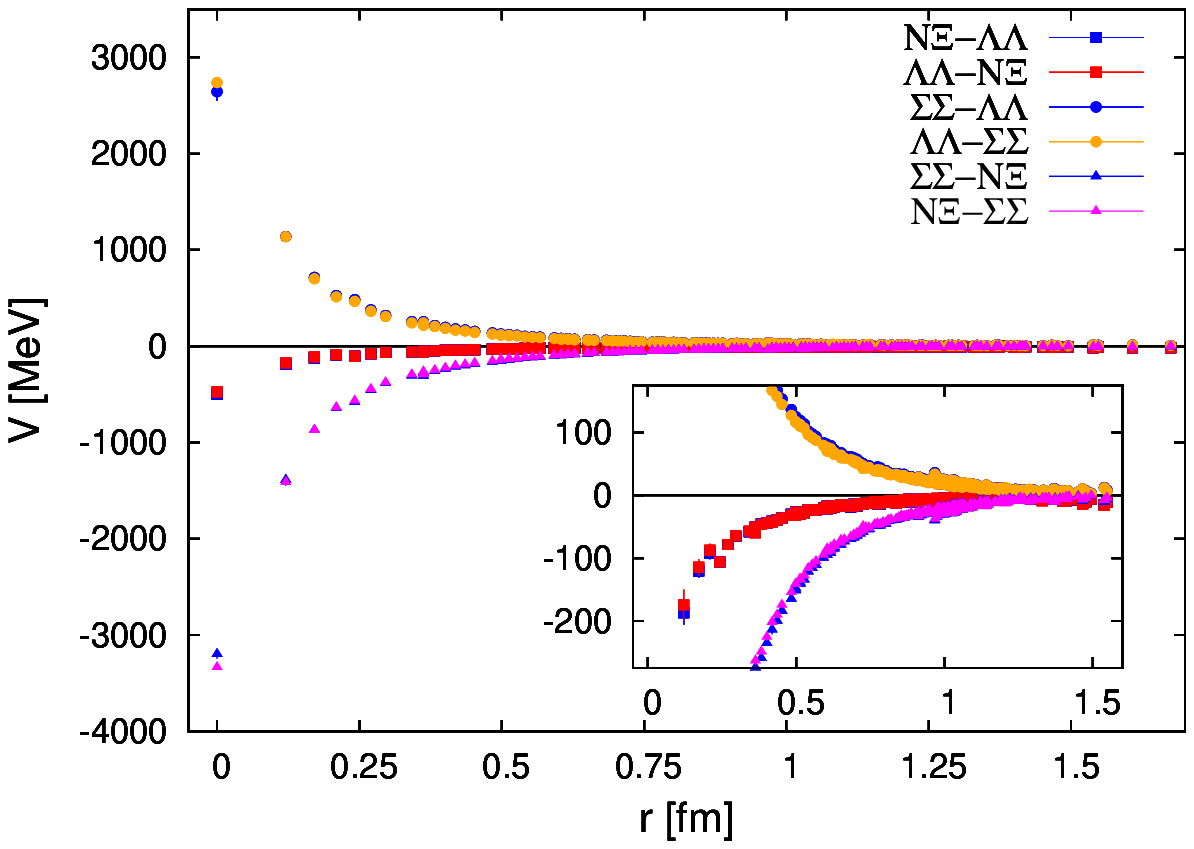}
\vspace*{-3mm}
\caption{
\label{fig:S=-2:coupled}
The coupled channel potentials 
of the $\Lambda\Lambda-N\Xi-\Sigma\Sigma$ system
at ($m_\pi$, $m_K$) $\simeq$ (875, 916) MeV~\cite{Aoki:2012tk}.
(Left) The diagonal parts of the potential matrix.
(Right) The off-diagonal parts of the potential matrix.
}
\end{center}
\vspace*{-6mm}
\end{figure}

In the realistic world, SU(3) symmetry is broken due to the heavy strange quark mass,
and it is often necessary to study the coupled channel systems above the inelastic threshold.
The HAL QCD method can be extended so that
the potentials above the inelastic threshold are extracted.
For instance, in the case of $\Lambda\Lambda-N\Xi-\Sigma\Sigma$ system,
we consider nine NBS wave functions
$
\phi_{X}^{W_i}(\vec{r}) =
 \langle 0 | B_{X_1}(\vec{r},0) B_{X_2}(\vec{0},0) | W_i \rangle_{\rm in} 
$
with $X(=X_1X_2) =$ $\Lambda\Lambda$, $N\Xi$ or $\Sigma\Sigma$
and three different energies $W_i$ $(i=0,1,2)$.
We consider the coupled channel Schr\"odinger equation
with these NBS wave functions,
and extract the energy-independent (non-local) $3\times 3$ potential matrix.
For details, see Refs.~\cite{Aoki:2011gt,Aoki:2012tk,Aoki:E-indep:inelastic}.

We employ $N_f = 2+1$ dynamical clover fermion configurations
generated by CP-PACS/JLQCD Collaborations~\cite{Ishikawa:2007nn}
on a $16^3\times 32$ lattice
at $a \simeq 0.12$ fm.
We calculate three quark mass setup,
corresponding to $m_\pi$ = 0.66, 0.75, 0.88 GeV 
with s-quark mass fixed at roughly physical value.
We perform a systematic study for $S= -1, -2, -3, -4$ channels,
which cover all 2B systems composed of two octet baryons.
(For a single channel study for $\Lambda N$ and $\Sigma N$ potentials 
 on a finer and larger volume by PACS-CS configurations,  see Ref.~\cite{Aoki:2012tk}.)
In this report, we present the results in $S= -2$, $I=0$ channel,
i.e., $\Lambda\Lambda-N\Xi-\Sigma\Sigma$ system,
which is relevant to the $H$-dibaryon.

In Fig.~\ref{fig:S=-2:coupled}, 
we show the coupled channel potential matrix
obtained at 
($m_\pi$, $m_K$, $m_N$, $m_\Lambda$, $m_\Sigma$, $m_\Xi$)
= (875(1), 916(1), 1806(3), 1835(3), 1841(3), 1867(2)) MeV~\cite{Aoki:2012tk}.
All diagonal components of the potential matrix have a repulsion at short distance,
while the strength of the repulsion in each channel varies reflecting properties 
of its main component in the irreducible representation of SU(3)$_f$. 
In particular, the $\Sigma \Sigma$ potential has the strongest repulsive core of these three components.
Note also 
that  off-diagonal parts of the potential matrix 
 satisfy the hermiticity 
within statistical errors.
We fit the potentials and solve the 
coupled channel Schr\"odinger equation,
and find that the $H$-dibaryon is still bound in this setup.
The results at smaller quark masses with PACS-CS configurations~\cite{Aoki:2008sm}, however,
show that
$H$-dibaryon tends to become a resonance 
with its energy moving from $\Lambda\Lambda$ threshold toward
$N\Xi$ threshold in our setup.
In order to obtain the definite conclusion on the fate of the $H$-dibaryon,
it is necessary to further reduce the quark mass toward the physical point.

\vspace*{-2mm}
\section{Three-Nucleon Forces}
\label{sec:3NF}
\vspace*{-2mm}

Three-nucleon forces (3NF) are considered to play 
an important role in various phenomena, e.g.,
the binding energies of light nuclei,
deuteron-proton elastic scattering,
the properties of neutron-rich nuclei
and 
nuclear EoS at high density relevant to the physics of neutron stars.
Together with experimental studies~\cite{Sekiguchi:2011ku},
lattice QCD is the most desirable way to determine 3NF.

The HAL QCD method can be extended to determine 3NF
by considering the NBS wave functions of three-nucleon (3N).
We introduce the imaginary-time NBS wave function of the 3N,
$\psi_{3N}(\vec{r},\vec{\rho},t)$,
defined by the six-point correlator as
\begin{eqnarray}
\label{eq:6pt_3N}
G_{3N} (\vec{r},\vec{\rho},t-t_0) 
&\equiv& 
\frac{1}{L^3}
\sum_{\vec{R}}
\langle 0 |
          (N(\vec{x}_1) N(\vec{x}_2) N (\vec{x}_3))(t) \
\overline{(N'       N'        N')}(t_0)
| 0 \rangle , \\[-1mm]
\label{eq:NBS_3N}
\psi_{3N}(\vec{r},\vec{\rho},t-t_0) &\equiv& G_{3N} (\vec{r},\vec{\rho},t-t_0) / e^{-3m_N (t-t_0)}
\end{eqnarray}
where
$\vec{R} \equiv ( \vec{x}_1 + \vec{x}_2 + \vec{x}_3 )/3$,
$\vec{r} \equiv \vec{x}_1 - \vec{x}_2$, 
$\vec{\rho} \equiv \vec{x}_3 - (\vec{x}_1 + \vec{x}_2)/2$
are the Jacobi coordinates.
Under the non-relativistic approximation,
existence of energy-independent potential for 3N systems can be shown
in a manner analogous to Sec.~\ref{sec:formalism}~\cite{Doi:2011gq,Aoki:E-indep:inelastic},
and we can employ 
the time-dependent HAL QCD method to extract
the 3NF without relying on the ground state saturation.
To proceed, 
employing the derivative expansion of the potentials,
the NBS wave function can be converted to the potentials
through the following 
Schr\"odinger equation,
\begin{eqnarray}
%
\biggl[ 
- \frac{1}{2\mu_r} \nabla^2_{r} - \frac{1}{2\mu_\rho} \nabla^2_{\rho} 
+ \sum_{i<j} V_{2N} (\vec{r}_{ij})
+ V_{3NF} (\vec{r}, \vec{\rho})
\biggr] \psi_{3N}(\vec{r}, \vec{\rho},t)
= - \frac{\partial}{\partial t} \psi_{3N}(\vec{r}, \vec{\rho},t) , \ \ \ \ 
\label{eq:Sch_3N}
\end{eqnarray}
where
$V_{2N}(\vec{r}_{ij})$ with $\vec{r}_{ij} \equiv \vec{x}_i - \vec{x}_j$
denotes 2NF between $(i,j)$-pair,
$V_{3NF}(\vec{r},\vec{\rho})$ the 3NF,
$\mu_r = m_N/2$, $\mu_\rho = 2m_N/3$ the reduced masses.
If we calculate 
$\psi_{3N}(\vec{r}, \vec{\rho},t)$,
and if all $V_{2N}(\vec{r}_{ij})$ are obtained
by (separate) lattice calculations for genuine 2N systems,
we can extract $V_{3NF}(\vec{r},\vec{\rho})$ through Eq.~(\ref{eq:Sch_3N}).

In our first exploratory study of 3NF,
we consider the total 3N quantum numbers of $(I, J^P)=(1/2,1/2^+)$, the triton channel.
We also restrict the geometry of the 3N.
More specifically, we consider the ``linear setup''with $\vec{\rho}=\vec{0}$,
with which 3N are aligned linearly with equal spacings of 
$r_2 \equiv |\vec{r}|/2$.
In this setup,
the third nucleon is attached
to $(1,2)$-nucleon pair with only S-wave,
and 
the wave function is completely spanned by
only three bases, which can be labeled
by the quantum numbers of $(1,2)$-pair as
$^1S_0$, $^3S_1$, $^3D_1$.
In this way,
the Schr\"odinger equation
can be simplified to
the $3\times 3$ coupled channel equations
with the bases of 
$\psi_{^1S_0}$, $\psi_{^3S_1}$, $\psi_{^3D_1}$.

% ...

While the computational cost of the NBS wave function is enormous,
it is drastically reduced (by a factor of 192)
using the ``unified contraction algorithm'' in Sec.~\ref{subsec:cost}~\cite{Doi:2012xd}.
We here employ the non-relativistic limit operator for the nucleon at the source, 
$N' = \epsilon_{abc}(q_a^T C \gamma_5 P_{nr} q_b) P_{nr} q_c$ with $P_{nr} = (1+\gamma_4)/2$,
to maximize the gain by the unified contraction algorithm.
For the nucleon operator at the sink, which defines the NBS wave function,
we employ the standard nucleon operator as in 2NF study,
$N = \epsilon_{abc}(q_a^T C \gamma_5 q_b) q_c$,
so that 2NF and 3NF are determined on the same footing.

In order to extract the genuine 3NF,
it is generally necessary to subtract 
the contributions from both of parity-even and parity-odd 2NF
from the total potential in the 3N system.
However, parity-odd 2NF on the lattice generally 
suffer from larger statistical errors due to the momentum injection in the system,
which could obscure the signal of 3NF.
Therefore, 
we consider the following channel,
%
%
%
%
$
\psi_S \equiv
\frac{1}{\sqrt{6}}
\Big[
-   \Pu \Nu \Nd + \Pu \Nd \Nu               
                - \Nu \Nd \Pu + \Nd \Nu \Pu 
+   \Nu \Pu \Nd               - \Nd \Pu \Nu
\Big]  ,
$
%
%
%
which is anti-symmetric
in spin/isospin spaces 
for any 2N-pair.
Combined with the Pauli-principle,
it is 
guaranteed that
any 2N-pair couples with even parity only,
and we can extract 3NF 
without referring to parity-odd 2NF.

Numerical simulations are performed by 
employing
$N_f=2$ clover fermion configurations
generated by CP-PACS Collaboration~\cite{Ali_Khan:2001tx},
at the lattice spacing of $a^{-1} = 1.269(14)$ GeV, 
the lattice size of $V = L^3 \times T = 16^3\times 32$,
and a quark mass corresponding to
$m_\pi = 1.1$ GeV and
$m_N = 2.2$ GeV.
They are the same configurations employed in Sec.~\ref{subsec:2NF:Podd}.

\begin{figure}[t]
\vspace*{-6mm}
\begin{minipage}{0.48\textwidth}
\begin{center}
\includegraphics[width=0.9\textwidth]{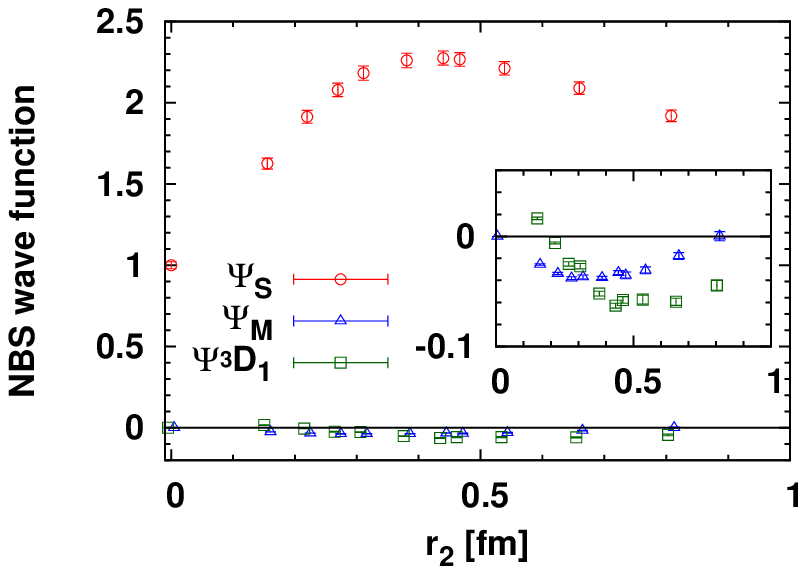}
\vspace*{-2mm}
\caption{
\label{fig:wf}
Obtained 3N wave functions, where
red, blue, green points are 
$\psi_S$, $\psi_M$, $\psi_{\,^3\!D_1}$, respectively.
}
\end{center}
\end{minipage}
\hfill
\begin{minipage}{0.48\textwidth}
\begin{center}
\includegraphics[width=0.9\textwidth]{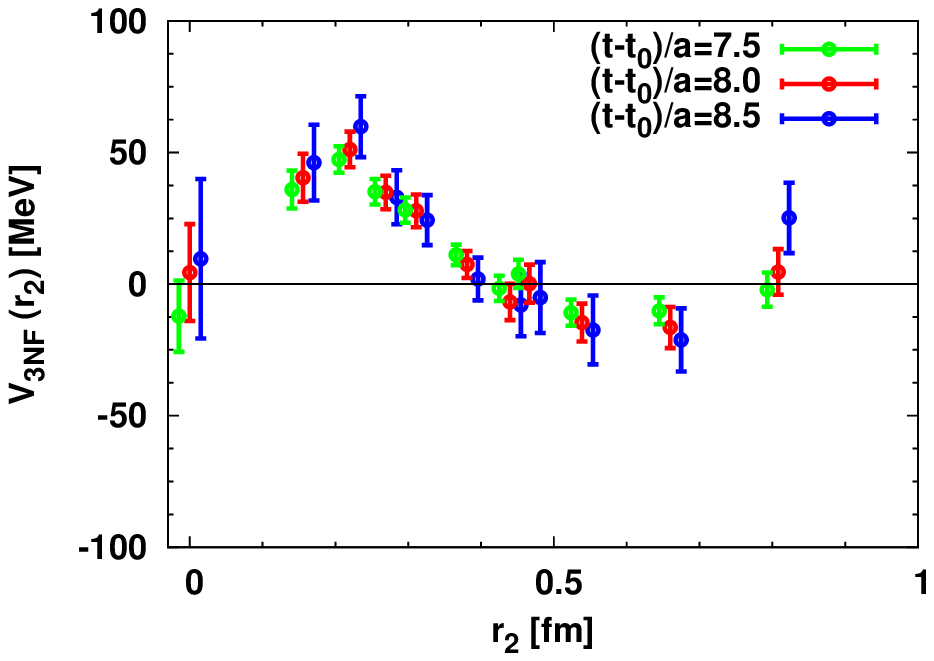}
\vspace*{-2mm}
\caption{
\label{fig:3N}
%
The genuine 3NF 
in the triton channel with the linear setup,
determined at each sink time.
%
}
\end{center}
\end{minipage}
\vspace*{-3mm}
\end{figure}

In Fig.~\ref{fig:wf},
we plot
the radial part of each wave function of
$\psi_S = ( - \psi_{^1S_0} + \psi_{^3S_1} )/\sqrt{2}$,
$\psi_M \equiv ( \psi_{^1S_0} + \psi_{^3S_1} )/\sqrt{2}$
and
$\psi_{^3D_1}$ 
obtained at $(t-t_0)/a = 8$,
which are 
normalized
by the central value of $\psi_S(r_2=0)$.
%
%
%
%
%
In Fig.~\ref{fig:3N}, we plot the preliminary results
for the genuine 3NF obtained at $(t-t_0)/a$ = 7.5, 8.0, 8.5,
where
results from different sink times are found to be consistent with each other.
Here, 3NF are effectively represented in a scalar-isoscalar functional form,
as is
often employed for
phenomenological short-range 3NF.
These results correspond to the update of those in Ref.~\cite{Doi:2011gq},
where the method is improved from original (time-independent) HAL QCD method 
to time-dependent 
one,
so that systematic errors associated with excited states are suppressed.

%
%
%
%
In Fig.~\ref{fig:3N},
an indication of repulsive 3NF is observed
at the short distance,
while 
3NF are found to be small at the long distance,
in accordance with the suppression
of two-pion exchange (2$\pi$E) 3NF by the heavy pion.
Note that a repulsive short-range 3NF
is phenomenologically required 
to explain the properties of high density matter.
%
%
%
The origin of the short-range 3NF may be attributed to the 
quark and gluon dynamics directly.
As discussed in Sec.~\ref{subsec:2BF:su3},
the short-range cores in 2B forces 
are well explained by the quark Pauli exclusion. 
In this context, 
it is intuitive to expect that the 3N system is subject to extra Pauli repulsion effect,
which could be an origin of the observed short-range repulsive 3NF.
%
It is also of interest that 
the analyses with operator product expansion~\cite{Aoki:OPE} show that
3NF has a universal repulsive core at short distance.

\vspace*{-3mm}
\section{Approaches from the L\"uscher's method}
\label{sec:luscher}

\begin{figure}[t]
\vspace*{-5mm}
\begin{center}
 \includegraphics[width=0.49\textwidth]{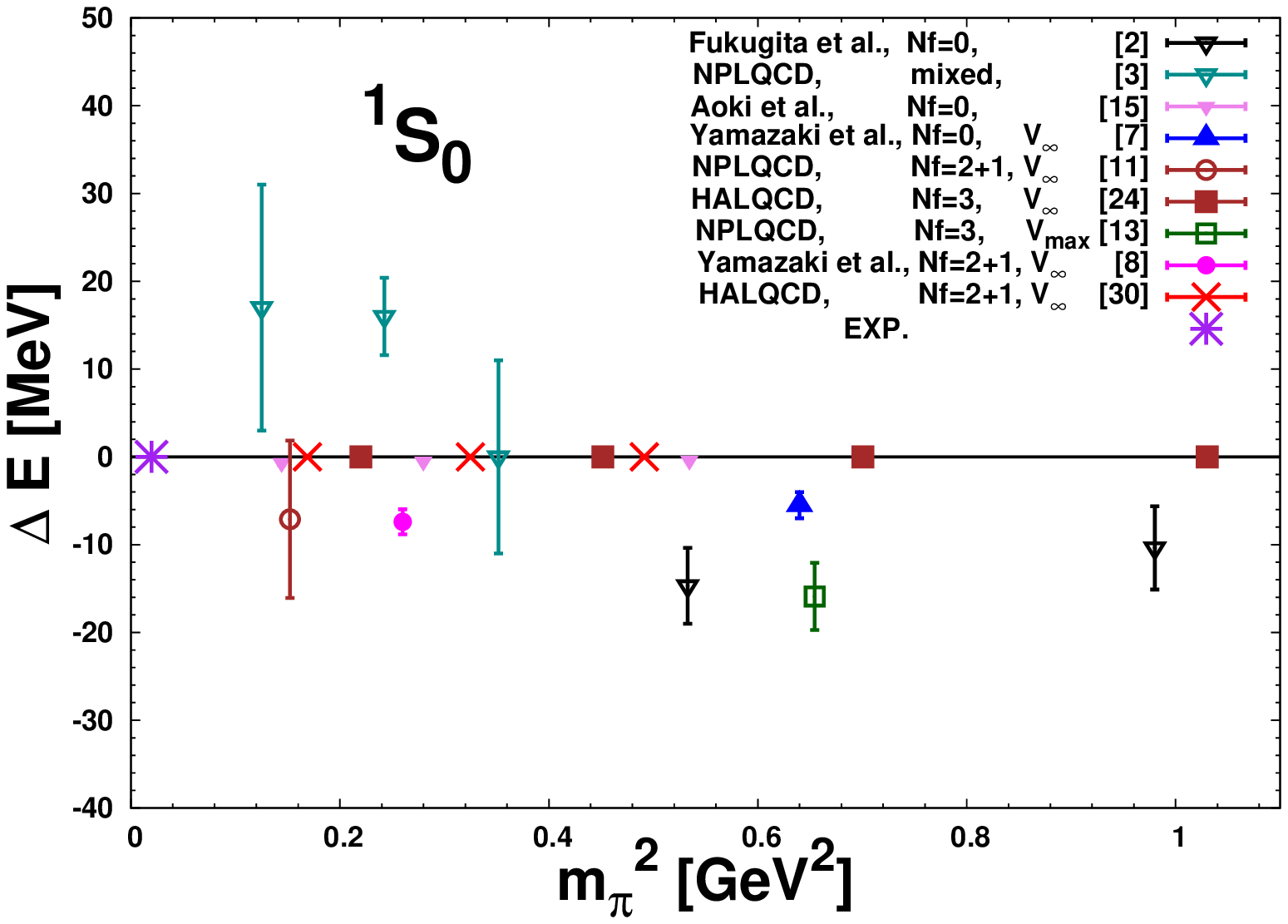}
 \includegraphics[width=0.49\textwidth]{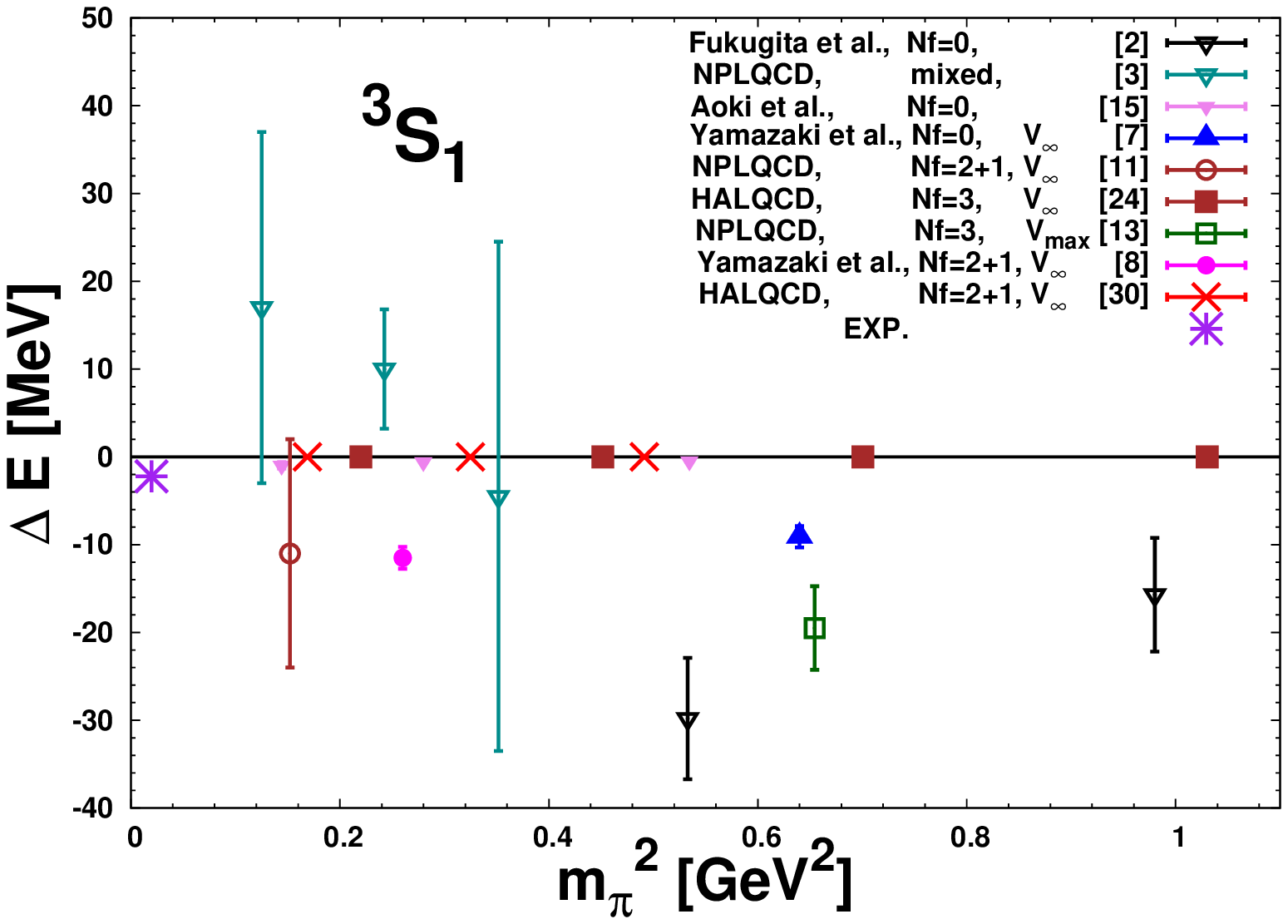}\hfill
 \includegraphics[width=0.49\textwidth]{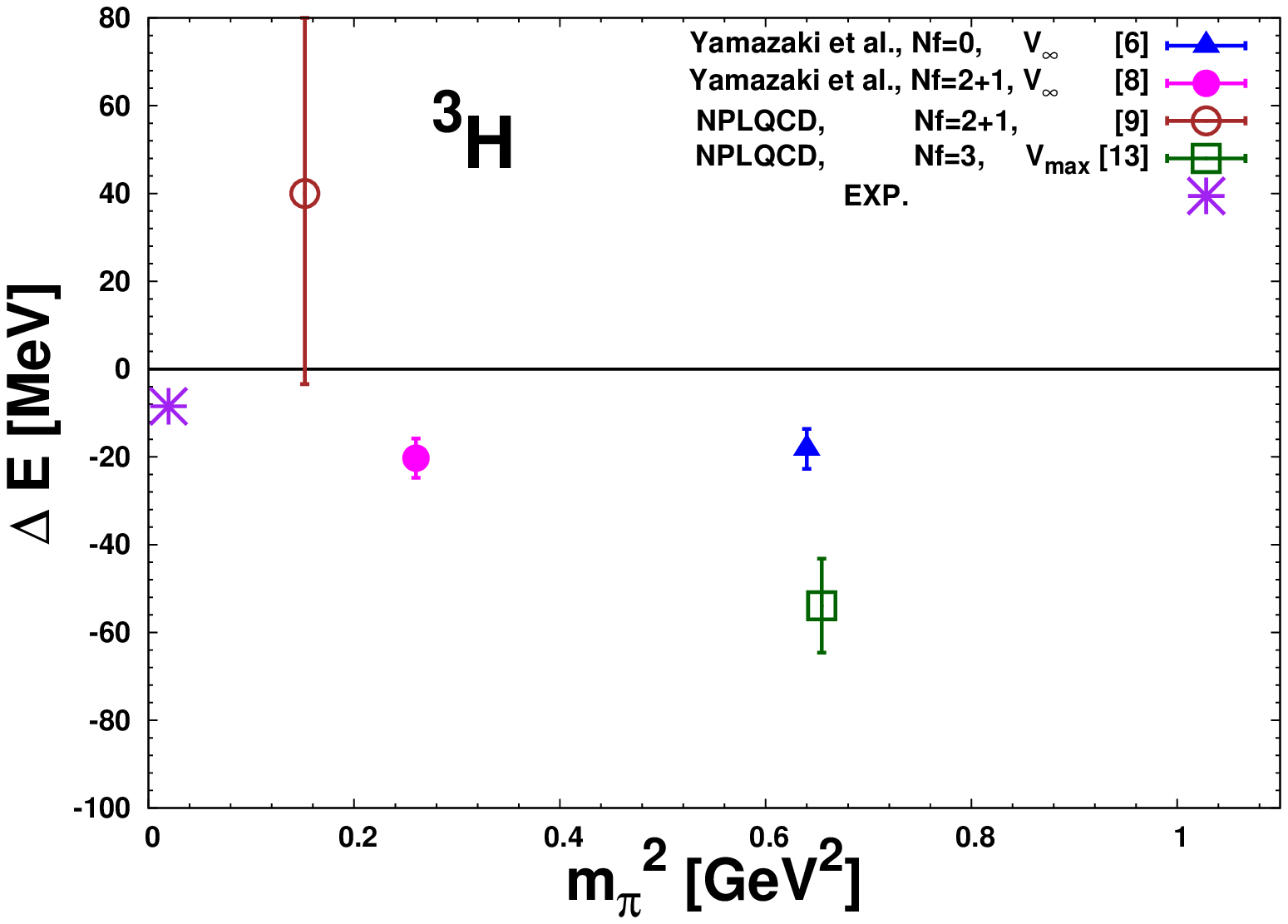}
 \includegraphics[width=0.49\textwidth]{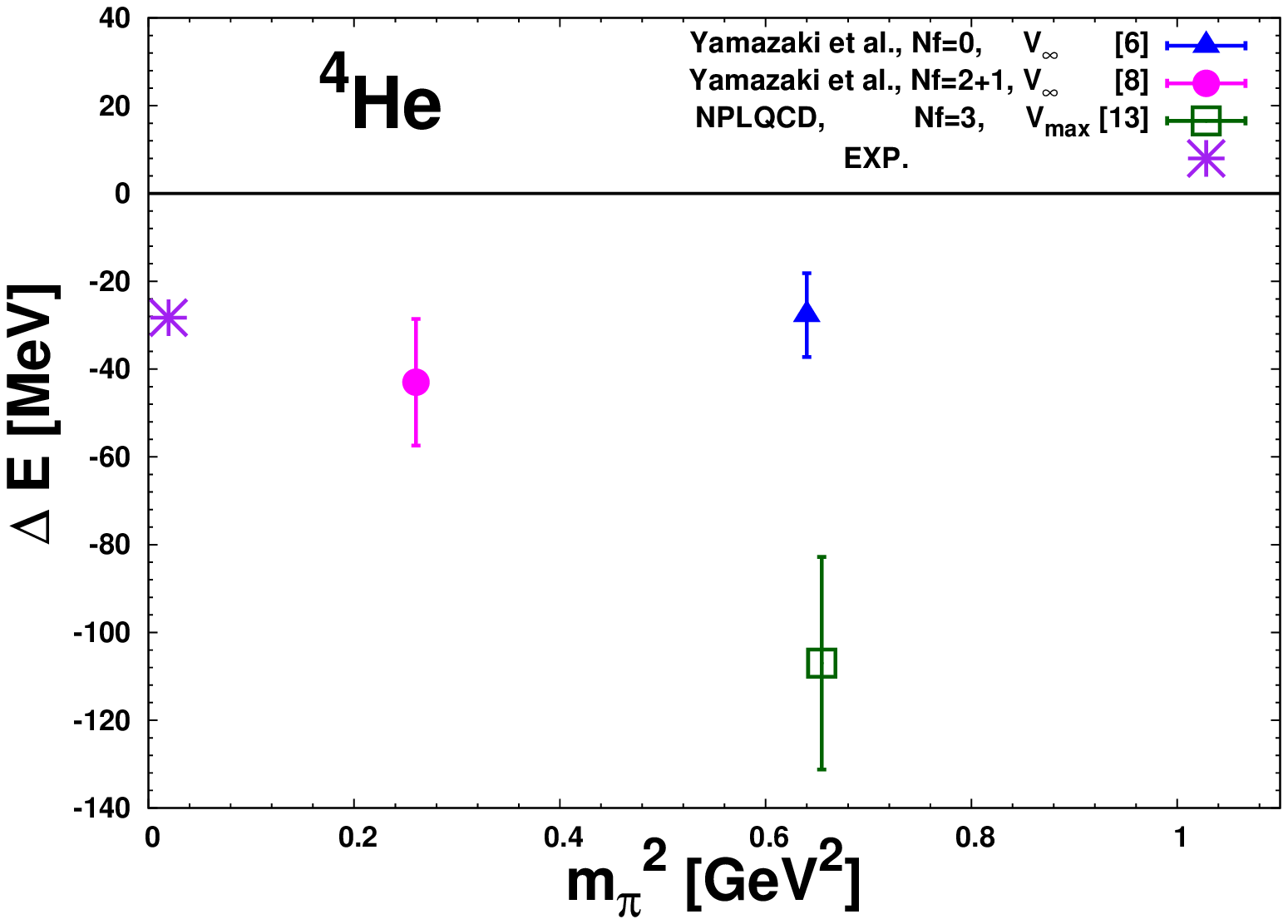}
\vspace*{-3mm}
\caption{
\label{fig:comp}
Compilation of $\Delta E$
in terms of $m_\pi^2$,
for
dineutron ($^1S_0$) (upper left),
deuteron ($^3S_1(-^3D_1)$) (upper right),
$^3$H  (=$^3$He) (lower left) and
$^4$He (lower right) channels.
Note that some results do not take $V\rightarrow \infty$.
}
\end{center}
\vspace*{-5mm}
\end{figure}

In the last few years,
a number of lattice studies 
have been performed
by extracting the energy of the system from 
the temporal correlator,
together with the L\"uscher's finite volume method. 

%
%
Yamazaki et al.~\cite{Yamazaki:2011nd}
performed quenched simulations 
with the clover fermion
at
$a = 0.128$ fm, $(m_\pi, m_N)$ = (0.80, 1.62) GeV
with the spacial lattice size of
$L = 3.1, 6.1, 12.3$ fm.
Single-state 
analyses were performed
with two quark smearing parameters. 
After the infinite volume extrapolation,
they observed that both of dineutron and deuteron 
are bound.
They also studied
$2\times 2$ diagonalization method
with $L = 4.1, 6.1$ fm.
The result for the 1st excited state is consistent with
the existence of a bound state, where 
the 2N operator was chosen so that the ground state energy
is consistent with single-state analysis.
They also performed the study of helium nuclei, 
using single-state analysis with two smearing parameters.
Both of $^3$He (=$^3$H) and $^4$He are found to be bound~\cite{Yamazaki:2009ua}.
Recently, 
they repeated the study with 
$N_f = 2+1$ full QCD simulations,
with clover fermion
at $a = 0.09$ fm, 
$(m_\pi, m_N)$ = (0.51, 1.32) GeV
with $L$ = 2.9--5.8 fm.
Single-state analysis shows that
all of dineutron, deuteron, $^3$He and $^4$He are bound~\cite{Yamazaki:2012hi}.



NPLQCD Collaboration performed $N_f = 2+1$ clover fermion simulations
on an anisotropic lattice 
at $a_s \sim$ 0.123 fm ($a_s/a_t \sim 3.5$),
$(m_\pi, m_N)$ = (0.39, 1.16) GeV 
with $L$ = 2.5 fm,
and found positive (repulsive) energy shifts in 
both of dineutron and deuteron channels~\cite{Beane:2009py},
as was observed in their mixed action study~\cite{Beane:2006mx}.
Hyperon interactions were studied as well,
and all are 
found to be repulsive except for $\Lambda\Lambda$ interaction~\cite{Beane:2009py}.
A feasibility study for three-baryon systems were also performed~\cite{Beane:2009gs,Beane:2010em}.
%
%
On the contrary,
their study including larger volumes ($L$ = 2.0, 2.5, 2.9, 3.9 fm)~\cite{Beane:2011iw}
found the suggestion of bound dineutron and deuteron
where only the results from largest two volumes were used.
They also found that $H$-dibaryon,
$\Xi^-\Xi^-$ and $n\Sigma^-(^1S_0)$ are bound~\cite{Beane:2010hg,Beane:2011iw,Beane:2012ey}.
%
%
Recently, they performed the simulation%
\footnote{
Their algorithm for the computation of correlators~\cite{Detmold:2012eu}
is nothing but the ``unified contraction algorithm'' which
was already proposed in Ref.~\cite{Doi:2012xd}.
}
in flavor SU(3) limit
with clover fermion 
at $a \sim 0.145$ fm (isotropic),
$(m_{\rm PS}, m_B)$ = (0.81, 1.64) GeV
with $L$ = 3.4, 4.5, 6.7 fm.
They obtained many bound (hyper-) nuclei,
including dineutron, deuteron,
flavor singlet $H$-dibaryon
and $^3$H, $^4$He nuclei~\cite{Beane:2012vq}.

The results of $\Delta E$,
which is the energy measured from the (fully-)breakup threshold,
for dineutron ($^1S_0$),
deuteron ($^3S_1-^3D_1$), $^3$H, $^4$He are summarized
in Fig.~\ref{fig:comp}.
While there exist differences in lattice setup,
we sometimes observe unexpected discrepancies between different results.
For instance, at $m_\pi \simeq 0.8$ GeV,
both of dineutron and deuteron are unbound in HAL QCD~\cite{Inoue:2011ai},
while both are deeply bound in NPLQCD~\cite{Beane:2012vq}.
The binding energy of flavor singlet $H$-dibaryon 
is also quite different between HAL QCD ($\Delta E = -37.8(3.1)(4.2)$ MeV)
and NPLQCD ($\Delta E = -74.6(3.3)(3.3)(0.8)$ MeV).
Since both studies employ $N_f = 3$ full QCD with a similar cut-off,
these discrepancies are open issues to be clarified,
which may be related to the difference in the analysis method.
The results from Yamazaki et al.~\cite{Yamazaki:2009ua,Yamazaki:2011nd} 
at $m_\pi = 0.8$ GeV are also quite different
from NPLQCD~\cite{Beane:2012vq}, although both groups employ basically the same analysis method.
It remains to be investigated whether the difference of simulation setup
(namely, $N_f = 0$ vs $N_f = 3$) can explain such large discrepancies.
Careful investigations on systematic errors should be examined,
e.g., excited state contaminations in the case of the 
traditional L\"uscher's method.
Note that, 
in the case of the time-dependent HAL QCD method,
the ground state saturation is no more required,
while the convergence of derivative expansion should be examined
for each channel of concern.

Finally, we note that
a study for the decuplet baryons has been also performed.
In Ref.~\cite{Buchoff:2012ja}, lattice simulations at $m_\pi = 0.39$ GeV
shows that $\Omega\Omega$ interaction in $J=0$ is weakly repulsive
with the scattering length of $a = -0.16(22)$ fm,
while $J=2$ is highly repulsive.
%
%

\vspace*{-2mm}
\section{Conclusions and Outlook}
\label{sec:summary}
\vspace*{-2mm}

We have presented lattice QCD activities for nuclear physics,
particularly the progress toward the 
determination of baryonic forces
using Nambu-Bethe-Salpeter wave functions.
Major challenges for multi-baryon systems on the lattice
have been addressed,
(i)  signal to noise (S/N) issue and
(ii) computational cost issue.
Recent breakthroughs on these issues have been given:
The S/N issue has been found to be avoided by the time-dependent HAL QCD method, 
in which 
energy-independent (non-local) potentials can be extracted 
without relying on the ground state saturation.
For the latter issue, 
a novel ``unified contraction algorithm'' has been developed,
by which computational cost is drastically reduced.
%
The lattice QCD results for 
nuclear forces, hyperon forces and three-nucleon forces have been presented,
and physical insights such as the origin of repulsive core have been discussed.
We have also shown 
recent results 
from the traditional L\"uscher's method,
and open issues to be resolved have been addressed.
Since the current simulations employ rather heavy quark masses,
it is crucial to go to lighter quark masses.
While there may appear various challenges toward the physical point simulations~\cite{field5},
it is becoming within reach to determine realistic nuclear forces including few-baryon forces 
from first-principles lattice simulations,
which will play an ultimate role in nuclear physics and astrophysics. 

\vspace*{-1mm}
\section*{Acknowledgments}
\vspace*{-2mm}

We thank authors and maintainers of CPS++\cite{CPS}.
We also thank  
the
PACS-CS,
CP-PACS
and 
JLQCD Collaborations
and ILDG/JLDG~\cite{conf:ildg/jldg} for providing gauge configurations.
The numerical simulations have been performed on 
Blue Gene/L, Blue Gene/Q and SR16000 at KEK,
SR16000 at YITP in Kyoto University, 
PACS-CS and T2K at University of Tsukuba,
and
T2K and FX10 at Tokyo University.
This research is supported in part by 
MEXT Grant-in-Aid for Young Scientists (B) (24740146),
Scientific Research on Innovative Areas (No.2004: 20105001, 20105003),
the Large Scale Simulation Program of KEK, 
the collaborative interdisciplinary program at T2K-Tsukuba, 
and SPIRE (Strategic Program for Innovative REsearch).
%


\end{document}